\documentclass[aps,pre,floatfix,showpacs,twocolumn, preprintnumbers]{revtex4-1}

\usepackage{caption}
\usepackage{subcaption}
\usepackage{graphicx}
\usepackage{amssymb}
\usepackage{amsmath}
\usepackage{mathtools}
\usepackage[dvipsnames,usenames]{xcolor}
\usepackage{multirow}
\usepackage{physics}
\usepackage{bm}

\begin{document}

\title{Solving the nuclear pairing model with neural network quantum states}
\author{ {Mauro} Rigo$^{\, {\rm a} }$,
{Benjamin} Hall$^{\, {\rm b,d} }$,
{Morten} Hjorth-Jensen$^{\, {\rm b,c} }$,
{Alessandro} Lovato$^{\, {\rm d, e, f} }$, and
{Francesco} Pederiva$^{\, {\rm a, f} }$
}

\affiliation{
$^{\,{\rm a}}$\mbox{Physics Department, University of Trento, via Sommarive 14, I-38123 Trento, Italy}\\
$^{\,{\rm b}}$\mbox{Department of Physics and Astronomy and Facility for Rare Ion Beams, Michigan State University, East Lansing, MI 48824, USA}\\
$^{\,{\rm c}}$\mbox{Department of Physics and Center for Computing in Science Education, University of Oslo, N-0316 Oslo, Norway}\\
$^{\,{\rm d}}$\mbox{Physics Division, Argonne National Laboratory, Argonne, IL 60439}\\
$^{\,{\rm e}}$\mbox{Computational Science Division, Argonne National Laboratory, Argonne, IL 60439}\\
$^{\,{\rm f}}$\mbox{INFN-TIFPA Trento Institute of Fundamental Physics and Applications, Via Sommarive, 14, 38123 {Trento}, Italy}\\
}
\date{\today}

\begin{abstract}
We present a variational Monte Carlo method that solves the nuclear many-body problem in the occupation number formalism exploiting an artificial neural network representation of the ground-state wave function. A memory-efficient version of the stochastic reconfiguration algorithm is developed to train the network by minimizing the expectation value of the Hamiltonian. We benchmark this approach against widely used nuclear many-body methods by solving a model used to describe pairing in nuclei for different types of interaction and different values of the interaction strength. Despite its polynomial computational cost, our method outperforms coupled-cluster and provides energies that are in excellent agreement with the numerically-exact full configuration interaction values. 
\end{abstract}

\maketitle

\section{Introduction}
The nuclear many-body problem entails non trivial challenges, primarily due to the non-perturbative nature and the strong spin-isospin dependence of realistic nuclear forces. Solving it has been the springboard for the development of sophisticated numerical methods that are designed to capitalize on each generation of high-performance computing resources~\cite{Hagen:2013nca,Hergert:2015awm,Soma:2020xhv,Lee:2020meg}. Among them, continuum quantum Monte Carlo approaches have proven extremely accurate in modeling nuclear dynamics at short and long range, but are either limited to relatively small systems or to somewhat simplified nuclear potentials~\cite{Carlson:2014vla}. On the other hand, although methods based on single-particle basis expansions can treat nuclei up to $^{208}$Pb --- including their binding energies, radii, and electroweak transitions --- in terms of the individual interactions among their constituents~\cite{Gysbers:2019uyb,Malbrunot-Ettenauer:2021fnr,Hu:2021trw}, they are not ideally suited to model the high-momentum components of the nuclear wave function. Therefore, a comprehensive description of short- and long-range nuclear dynamics for medium-mass and heavy nuclei has not been achieved using available nuclear many-body methods. 

Since their pioneering application to interacting spin models~\cite{Carleo:2017}, neural network quantum states (NQS) have seen widespread and successful applications to quantum chemistry~\cite{Choo:2019,Hermann:2019,Pfau:2019} and condensed matter~\cite{Pescia:2021kxb,Wilson:2022meh} problems. When solving many-particle systems of fermions, NQS must respect the Pauli exclusion principle; in the first quantization formalism, this constraint is imposed by using representations that are anti-symmetric by construction, such as the Slater-Jastrow~\cite{Stokes:2020ihf}, the generalized backflow~\cite{Hermann:2019,Pfau:2019}, and the hidden-fermion~\cite{Moreno:2021jas} ansatzes. On the other hand, within the second quantization formalism, the Pauli principle is automatically taken care of, thereby allowing one to use simpler NQS architectures, including restricted Boltzmann Machines~\cite{Choo:2019,Zhao:2022}. 

Within low-energy nuclear Physics, from their initial applications to solving the deuteron in momentum space~\cite{Keeble:2019bkv,Sarmiento:2022bxn}, NQS have been subsequently combined with Variational Monte Carlo (VMC) techniques to approximately solve $A\leq 4$~\cite{Adams:2020aax} and $A\leq 6$ nuclei~\cite{Gnech:2021wfn}. Most recently, the hidden-nucleon architecture has been proposed to overcome the limitations of the Slater-Jastrow ansatz and successfully applied to nuclei up to $^{16}$O~\cite{Lovato:2022tjh}. Nuclear Physics applications to NQS have so far been limited to the first-quantization formalism. However, the possibility of using non-diverging effective nuclear forces at short distances makes NQS in the occupation-number formalism ideally suited to tackle the nuclear many-body problem. 

In this work, we put forward a novel VMC method suitable for solving the nuclear many-body problem using NQS to compactly represent the ground state of the system in the occupation-number formalism. As a first application, we consider different versions of the nuclear pairing model, which is one of the most important ingredients of the effective nuclear interaction, as recognized in the early work of Bohr, Mottelson, and Pines~\cite{Bohr:1958zz}. Pairing plays an important role in the calculations of a number of properties that depend on the low-energy microscopic structure of the nucleus, including their ground-state energies, low-lying excitations, level densities, odd-even staggering effects, single-particle occupancies, electromagnetic transition rates. Pairing is also relevant in our understanding of nuclear reactions and particle decays~\cite{Dean:2002zx,Moller:1992zz}.

Solving the pairing Hamiltonian, even in the simple case of constant pairing strength, involves nontrivial difficulties. Exact diagonalization techniques~\cite{Molique:1997zz,Volya:2000ne,Zelevinsky:2003ad,Liu:2020mkp} and particle-number conserved seniority approaches~\cite{Sandulescu:1996wq,Jia:2013lea} suffer from growing exponential
complexity and are therefore very hard to use for
large systems. However, different families of integrable pairing models were recognized after the pioneering work of Richardson~\cite{Richardson:1963a,Richardson:1963b}, and are now known as Richardson-Gaudin models~\cite{Gaudin:1976}. As an example of their physical relevance, the authors of Ref.~\cite{Dukelsky:2011nb} found that a separable pairing interaction and non-degenerate single-particle energies with two free parameters provides an excellent approximation to the Gogny pairing. It has to be noted that the energies of the Richardson-Gaudin models are found by solving a set of coupled non-linear equations, which by itself is a complicated problem~\cite{Guan:2022CoPhC} and it is still an area of active research. 

In this work, we compare the VMC-NQS method with many-body perturbation theory, coupled-cluster calculations, exact-diagonalization techniques, and iterative solutions of the Richardson-Gaudin equations. We consider different values of the interaction strength, and both constant and separable couplings. 

This work is organized as follows. In Section~\ref{sec:pairing} we discuss the different versions of the pairing model employed in this work and their exact solutions. In Section~\ref{sec:VMC} we introduce the VMC method based on NQS. Section~\ref{sec:coupled_cluster} is dedicated to coupled-cluster theory, with particular emphasis to pair coupled cluster doubles theory. Section~\ref{sec:manybody} describes a many-body perturbation theory approach. In Section~\ref{sec:results} we present our results and 
concluding remarks are given in Section~\ref{sec:conclusions}.

\section{Nuclear pairing model}
\label{sec:pairing}

\subsection{The Model}

\begin{figure}
    \centering
    \includegraphics[width=0.25\textwidth]{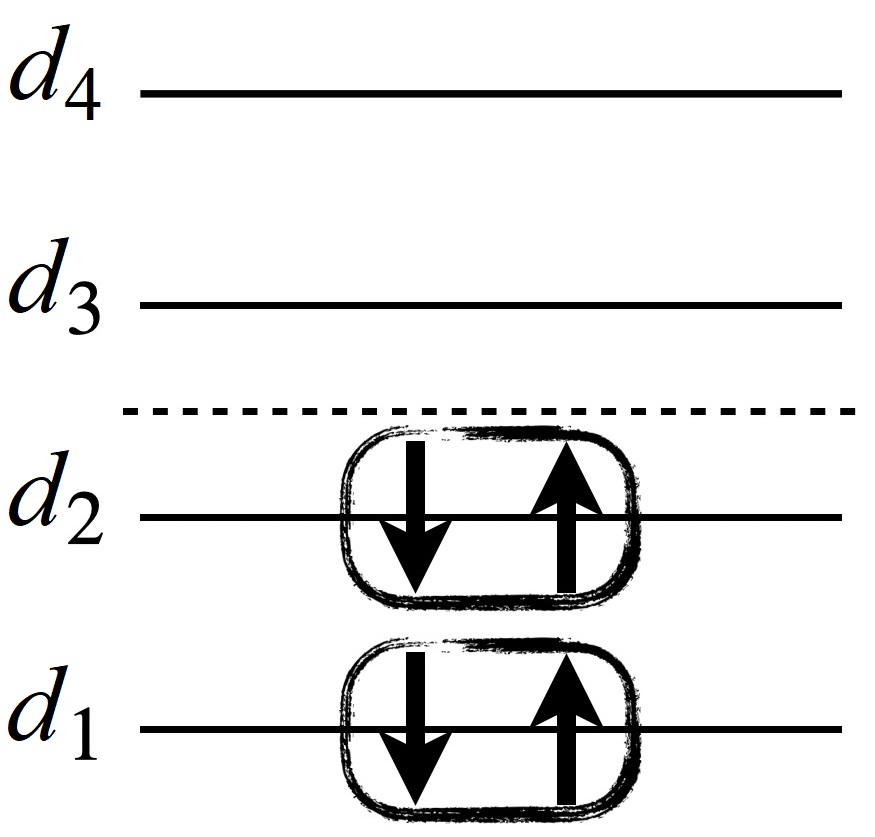}
    \caption{Example of model space with $P=4$ and $M=2$. Shown are four energy levels with single-particle energies $d_1,\,d_2,\,d_3,\,d_4$, of which the bottom two are initially filled by pairs. The dashed line represents the Fermi level which divides the energy levels with energies $d_1$ and $d_2$ (the hole states) from those with energies $d_3$ and $d_4$ (the particle states).}
    \label{fig:pairing_schem}
\end{figure}

In 1963, R.W. Richardson proposed a model consisting of fermions occupying non-degenerate energy levels which interact solely through the pairing force~\cite{Richardson:1963a,Richardson:1963b}. This pairing model consists of $P$ non-degenerate energy levels, occupied by $M$ pairs of fermions of opposite spin. Its Hamiltonian is given by
\begin{align}
\label{pairing_model_hamiltonian_original}
H=\sum_{p\sigma}d_pa^\dagger_{p\sigma}a_{p\sigma}-\sum_{pq}g_{pq}a^{\dagger}_{p+}a^{\dagger}_{p-}a_{q-}a_{q+}\ ,
\end{align}
where the indices $p$ and $q$ sum over the set $\{1,...,P\}$, representing the different energy levels, while $\sigma$ runs over the set $\{+,-\}$, corresponding to the spin of each fermion. The coefficients $d_p$ denote the single-particle energies, which for our our analysis we set to grow linearly as $d_p=p$. The coefficients $g_{pq}$ are the so-called pairing strengths, representing the energy associated with moving a pair of fermions from the $q^{\text{th}}$ to the $p^{\text{th}}$ energy level. 

The pairing Hamiltonian can be rewritten in terms of so-called pairing operators as
\begin{align}
\label{pairing_model_hamiltonian}
H=\sum_{p=1}^Pd_pN_p-\sum_{p,q=1}^Pg_{pq}A^\dagger_pA_q\ .
\end{align}
Here, $N_p$ is the pair number operator, which counts the number of fermions occupying the $p^{\text{th}}$ energy level. Furthermore, $A^\dagger_p$ and $A_p$ are the pair fermionic creation and annihilation operators respectively, which create and annihilate pairs of fermions on the $p^{\text{th}}$ energy level. These operators are defined in terms of fermionic creation and annihilation operators as follows:
\begin{align}
N_p &= \sum_{\sigma}a^\dagger_{p\sigma}a_{p\sigma}\nonumber
\\
A^{\dagger}_p &= a^{\dagger}_{p+}a^{\dagger}_{p-}\label{pair_fermionic_operators}
\\
A_p &= a_{p-}a_{p+}\nonumber
\end{align}
where $\sigma$, again, sums over the set $\{+,-\}$. The purpose of this rewriting becomes clear when one notices that these operators satisfy the $\text{SU}(2)$ algebra described by the following commutation relations:
\begin{align}
[A_p,A^\dagger_q]&=\delta_{pq}(1-N_p)\nonumber
\\
[N_p,A^\dagger_q]&=2\delta_{pq}A^\dagger_p
\\
[N_p,A_q]&=-2\delta_{pq}A_p\nonumber
\end{align}

\subsection{Pairings and exact solutions}

When Richardson originally introduced the pairing model, he accompanied it with an exact solution for the case of constant pairing strengths, $g_{pq}=g$ for all $p$ and $q$,
\begin{align}
\label{pairing_model_hamiltonian_constant_g}
H=\sum_{p=1}^{P}d_pN_p-g\sum_{p,q=1}^{P}A^\dagger_pA_q\ .
\end{align}
The ansatz that solves the model with $M$ pairs is
\begin{align}
\label{richardson_ansatz}
\ket{\Psi}=\prod_{\alpha=1}^M B^\dagger_{\alpha}\ket{0}\ ,
\end{align}
where
\begin{align}
B^\dagger_\alpha=\sum_{\kappa=1}^P\frac{1}{2d_\kappa-E_\alpha}A^\dagger_\kappa\ .
\end{align}
When plugged into the Schr\"odinger equation, this solution leads to the so-called Richardson equations~\cite{Richardson:1963a,Richardson:1963b}
\begin{equation}
\label{richardson_equations}
1 - \sum_{\kappa=1}^P \frac{g}{2d_\kappa-E_\alpha}
- \sum_{\beta=1,\beta\neq\alpha}^M \frac{2g}{E_\alpha-E_\beta}
= 0\ .
\end{equation}
This is a set of coupled, non-linear equations which one solves for the terms $E_\alpha$ and the ground-state energy of the Hamiltonian is found as $E=\sum_{\alpha=1}^M E_\alpha$. Unfortunately, the algebraic solution of the Richardson equations becomes numerically unstable at certain critical values of the interaction strength, and dedicated stable algorithms have been developed to this aim~\cite{Rombouts:2003zd,Guan:2022CoPhC}.

In addition to the constant-pairing Hamiltonian above, in this work we also consider the exactly solvable Hamiltonian with separable pairing interaction and non-degenerate single-particle energies 
\begin{align}
\label{pairing_model_hamiltonian_separable_g}
H=\sum_{p=1}^Pd_pN_p- 2 g \sum_{p,q=1}^P\sqrt{(\alpha-d_p)(\alpha-d_q)} A^\dagger_pA_q
\end{align}
where $g$ is the pairing strength and $\alpha$ the interaction cutoff. The above Hamiltonian belongs to the hyperbolic family of Richardson-Gaudin models, whose exact solution can be obtained by again solving a set of coupled non-linear equations similar to Eq.~\eqref{richardson_equations}. Despite its simplicity, this model is capable of reproducing properties of heavy nuclei as described by a more realistic Gogny interaction~\cite{Dukelsky:2011nb}.

\section{Variational Monte Carlo with neural quantum states}\label{sec:VMC}
Variational Monte Carlo (VMC) algorithms solve the many-body Schr\"odinger equation of a given Hamiltonian $H$ by approximating the true ground state of the system with a variational state $\ket{\psi_V}$, defined in terms of a set of variational parameters $\bm \theta$. Their optimal values are found by minimizing the variational energy
\begin{equation}
E_V(\bm\theta) \equiv \frac{\bra{\psi_V}H\ket{\psi_V}}{\braket{\psi_V}}\ ,
\label{eq:variational}
\end{equation}
since the Rayleigh–Ritz variational principle ensures that the true ground-state energy $E_0\leq E_V(\bm \theta)$. Therefore, a minimal $E_V(\bm\theta)$ corresponds to the best possible approximation to the true ground-state wave function given the ansatz $\ket{\psi_V}$.

In this work, we expand the variational wave function in the occupation-number basis, where each state $\ket{N} = \ket{n_1,\,n_2,\,...,\,n_P}$ is characterized by a set of occupation numbers $n_p=0,\,1$ representing the number of pairs occupying each energy level $p=1,\,...,\,P$. By inserting a completeness relation over the occupation states, $1 = \sum_N|N\rangle\langle N|$, in Eq.~\eqref{eq:variational}, one obtains
\begin{align}
\label{vmc_avg}
    E_V(\bm\theta)=\sum_{N}\frac{\vert\bra{\psi_V}N\rangle\vert^2}{\braket{\psi_V}}E_L(N)\ ,
\end{align}
where we define the local energy as
\begin{equation}
E_L(N) = \frac{\bra{N}H\ket{\psi_V}}{\bra{N}\psi_V\rangle}\ .
\end{equation}
The sum entering Eq.~\eqref{vmc_avg} quickly becomes intractable by means of exact-quadrature methods due to the high the dimensionality of the Hilbert space, and it is typically evaluated by means of Monte Carlo methods. Indeed, Eq.~\eqref{vmc_avg} has already the form of an expectation value of the local energy
over the normalized probability distribution
\begin{equation}
P(N)=\frac{\vert\bra{\psi_V}N\rangle\vert^2}{\braket{\psi_V}}\ .
\label{eq:prob_N}
\end{equation}
Therefore, according to the central limit theorem, $E_V(\bm \theta)$ can be estimated by the average of the local energy $E_L(N)$ calculated for a set of configurations $N$ sampled according to the probability $P(N)$. This latter task can be accomplished by employing the Metropolis-Hastings algorithm~\cite{Metropolis:1953,Hastings:1970}. In order for the exclusion principle to be satisfied and the total number of pairs $\sum_p n_p$ to be conserved, in the random walk, the proposed state $\ket{N^\prime}$ is obtained by swapping two random occupation numbers in the array of 0s and 1s defining a given $\ket{N}$. Note that this procedure is not limited to the Hamiltonian operator, but it can be used to estimate the ground-state expectation value of any quantum-mechanical operator.

When evaluating $E_L(N)$, the one-body term of the Hamiltonian and the diagonal part of the pairing interaction can be computed as
\begin{align}
&\bra{N}\sum_p\Big(d_pN_p - g_{pp}A^\dag_pA_p\Big)\ket{\psi_V} \nonumber\\
&\qquad =\sum_p(2d_p-g_{p p})n_p \bra{N}\psi_V\rangle\ .
\end{align}
To compute the non-diagonal terms, we insert a completeness relation:
\begin{align}
    &\bra{N}\sum_{p\neq q}g_{pq}A^\dag_pA_q \ket{\psi_V}\nonumber\\
    & \qquad = \sum_{N'}\bra{N}\sum_{p\neq q}g_{pq}A^\dag_pA_q\ketbra{N'}\psi_V\rangle\ .
\end{align}
In this expression, for a given a state $\ket{N}$, only the indices $p,\,q$ such that $n_p=1$ and $n_q=0$ contribute to the inner sum. Furthermore, for fixed $p,\,q$, only the $\ket{N'}$ having $n'_p=0,\,n'_q=1$ and $n'_r=n_r$ for $r\neq p,\,q$ yields a non-zero matrix element. Thus, these terms can be calculated using two nested loops, one with index $p$ running over the non-zero occupation numbers of $\ket{N}$ and the other with index $q$ over the null ones, and multiplying $g_{pq}$ by the variational wave function evaluated for a state $\ket{N'}$ with the same occupation numbers as $\ket{N}$ but $n'_p=0,\,n'_q=1$.

\subsection{Neural network quantum states}
Inspired by the recently introduced NQS ansatzes~\cite{Gnech:2021wfn,Lovato:2022tjh}, we parametrize the variational wave function as
\begin{equation}
    \psi_V(N)\equiv \bra{N}\psi_V\rangle = e^{\mathcal{U}(N)}\tanh(\mathcal{V}(N))\, ,
\end{equation}
where $\mathcal{U}(N)$ and $\mathcal{V}(N)$ are fully-connected neural networks (FCNNs). In contrast with previous applications in the field based on complex variational states, our ansatz is real-valued; the function $\mathcal{U}(N)$ determines the magnitude of $\psi_V(N)$, while $\tanh(\mathcal{V}(N))$ is a smooth version of its sign. 

A chief advantage of using the occupation number basis is that the states $\ket{N}$ inherently encode the exchange symmetry required for a wave function describing a system of fermions, thus the FCNNs can simply take as inputs the string of occupation numbers $n_p=0,1$ defining each state. We typically use between $2$ to $3$ hidden layers and \textit{tanh} activation functions, since these choices proved to be stable and yield the most accurate results. As we typically use hidden layers of $\sim P$ neurons, evaluating the wave function exhibits a polynomial scaling in the number of energy levels ($\mathcal{O}(P^2)$), independent of the number of pairs. With a complexity for the calculation of the local energy at worst quadratic in the number of energy levels, this gives the whole algorithm a complexity in the number of energy levels of $\mathcal{O}(P^4)$.

\subsection{Training algorithm}\label{VMC:training}
A significant improvement in the execution time and in the stability of the optimization can be achieved by using the so-called Stochastic Reconfiguration (SR) algorithm \cite{Sorella:2005}, which is closely related to the natural gradient descent method~\cite{amari_natural_1998, Stokes:2019}. Within the SR algorithm, the variational parameters $\bm \theta$ are updated as
\begin{align}
\bm\theta\to\bm \theta+\delta \bm \theta \ \ ,\quad 
    \bm S\delta\bm \theta = -\delta\tau\bm G 
    \label{sr_update}
\end{align}
where $\delta\tau$ is the learning rate, $\bm S$ is the overlap matrix and $\bm G$ is the gradient of the variational energy with respect to the variational parameters. The matrix elements of $\bm S$ and $\bm G$ are calculated as the following covariances:
\begin{align}
    S_{mn} &= \langle \mathcal{O}_m^\dag\mathcal{O}_n\rangle - \langle \mathcal{O}_m^\dag\rangle\langle\mathcal{O}_n\rangle\ ,\label{SRmat}\\
    G_n &= 2\langle\mathcal{O}_n^\dag H\rangle - 2\langle \mathcal{O}_n^\dag\rangle\langle H\rangle\ ,\label{sr_grad}
\end{align}
where 
$\mathcal{O}_n\ket{\psi_V}=\partial\ket{\psi_V}/\partial\theta_n$ is the derivative operator with respect to the variational parameter $\theta_n$. An optimization using this update is essentially equivalent to performing an imaginary-time evolution of the variational wave function with time step $\delta\tau$~\cite{sorella_green_1998,Sorella:2005} in the manifold spanned by the wave function and its derivatives. Since the overlap matrix is positive definite, the linear system of Eq.~\eqref{sr_update} could be solved using, e.g., the Cholesky decomposition. However, this requires storing the entire matrix $\bm S$ and, since the number of variational parameters is of order $P^2$, its size scales as $P^4$, which quickly becomes prohibitively large as $P$ increases. For this reason, we solve the linear system using the conjugate gradient algorithm, which only requires the definition of a function that calculates the matrix-vector product $\bm S \bm v$ for an arbitrary vector $\bm v$. The expectation values in Eqs.~\eqref{SRmat} and~\eqref{sr_grad} can once again be expressed as averages over configurations sampled according to the probability $P(N)$ of Eq.~\eqref{eq:prob_N}; the product $\bm S \bm v$ given an arbitrary $\bm v$ is then rewritten in order to make its evaluation time and memory efficient~\cite{Neuscamman:2012}.
Following Ref.~\cite{Lovato:2022tjh}, to further improve the stability of the convergence, we regularize the SR equations by adding a small RMSProp-inspired diagonal shift as:
\begin{align}
    \bm v \to \beta \bm v + (1-\beta)\bm G^2\ \ ,&\nonumber\\
    \left[\bm S + \varepsilon\,\text{diag}\left(\sqrt{\frac{\bm v}{ (1-\beta^k)}}+10^{-8}\right)\right]\delta\bm\theta &= -\delta\tau\bm G\, .
    \label{sr_update_new}
\end{align}
In the above equation, $\varepsilon$ is the strength of the diagonal shift, whose typical values range from $\varepsilon=0.0001$ to $\varepsilon=0.001$ for different Hamiltonians. As with the exponential decay rates for the second moment estimates of the RMSprop algorithm~\cite{Hinton:2018}, we take $\beta=0.99$. Finally, the factor $(1-\beta^k)$ corrects the bias in the second-moment estimates as in Adam~\cite{Kingma:2014}, with $k$ being the current optimization step number. 

\section{Pair Coupled Cluster Doubles Theory}
\label{sec:coupled_cluster}
Coupled cluster theory was first developed to solve the nuclear Schr\"odinger equation by Coester and K\"{u}mmel in the late 1950's and early 1960's~\cite{Coester:1958,Coester:1960}. In this work, we focus on the pair coupled cluster doubles theory (pCCD), which is ideally suited to solve the pairing problem. The pCCD ansatz is given by
\begin{align}
\label{cc_ansatz}
\ket{\Psi}=e^{T}\ket{\Phi_0}
\end{align}
where $\ket{\Phi_0}$ is the reference state and $T$ is the cluster operator restricted to moving pairs of fermions:
\begin{align}
T\equiv\sum_{ia}t_{ai}A^\dagger_a A_i
\end{align}
Here, $A^\dagger_a$ and $A_i$ are the pair fermionic creation and annihilation operators defined in Eq.~\eqref{pair_fermionic_operators}. Inspired by the notation commonly adopted in quantum chemistry, we reserve the labels $i,j,k,\dots$ for the hole states and $a,b,c,\dots$ for the particle states.

Starting from the time-independent Schr\"{o}dinger equation \begin{align}
H\ket{\Psi}=E\ket{\Psi}
\end{align}
with the coupled cluster ansatz (\ref{cc_ansatz}), left-multiplying both sides by $e^{-T}$ and left-multiplying by either the reference state or an excited state yields
\begin{align}
\label{pCCD_energy}
E&=\mel{\Phi_0}{\overline{H}}{\Phi_0}\ ,
\\
\label{pCCD_eq}
0&=\mel{\Phi_i^a}{\overline{H}}{\Phi_0}\ ,
\end{align}
where the excited state $\bra{\Phi_i^a}$ is obtained as
\begin{equation}
\bra{\Phi_i^a}=\bra{\Phi_0}A^\dagger_iA_a
\end{equation}
and $\overline{H}$ is the similarity-transformed pairing Hamiltonian:
\begin{align}
\overline{H}=e^{-T}He^T\ .
\end{align}
Using the Baker–Campbell–Hausdorff (BCH) identity, $\overline{H}$ can be expanded as
\begin{align}
\label{sim_H_p}
\overline{H}=H+\comm{H}{T}+\frac{1}{2}\comm{\comm{H}{T}}{T}+\cdots
\end{align}
When the latter expression is plugged into the pCCD equations~\eqref{pCCD_energy} and~\eqref{pCCD_eq}, one can truncate the resulting expressions by noting that only certain terms in the infinite sum (those which can be fully contracted) are non-zero. Separating the pairing model Hamiltonian into a single-body and a two-body term $H=F+V$, where
\begin{align}
F &= \sum_pd_pN_p\ , \nonumber \\
V &= -\sum_{pq}g_{pq}A^\dagger_pA_q\ ,
\end{align}
the truncated pCCD equations read
\begin{align}
E &= \bra*{\Phi_0}F+V+V T\ket{\Phi_0}\ , \label{pCCD_energy_trunc} \\
0 &=
\bra*{\Phi_i^a}V+F T+V T +\frac{1}{2}V T^2-T V T \ket{\Phi_0}\label{trunc_pCCD_eq}\ .
\end{align}
Recognizing that the first two terms of the truncated pCCD energy in Eq.~\eqref{pCCD_energy_trunc} equal the reference energy, the correlation energy equation can readily be identified with
\begin{equation}
\label{pCCD_e_corr}
\Delta E = \mel{\Phi_0}{V T}{\Phi_0}=-\sum_{ia}t_{ai}g_{ia}\ ,
\end{equation}
where in the last equality we used Wick's theorem \cite{Wick1950}. Similarly, Eqs.~\eqref{trunc_pCCD_eq} become
\begin{align}
\label{pCCD_eq_exp}
0& =g_{ai}+2\Big{(}d_i-d_a-g_{ii}+g_{ia}t_{ai}-\sum_b g_{ib}t_{bi}\nonumber\\
&-\sum_jg_{ja}t_{aj}\Big{)}t_{ai}+\sum_bg_{ab}t_{bi}+\sum_jg_{ji}t_{aj}\nonumber\\
&+\sum_{bj}g_{jb}t_{aj}t_{bi}\ .
\end{align}
In practice, the amplitudes $t$ are found by solving the non-linear and coupled Eqs.~\eqref{pCCD_eq_exp}, which is accomplished using an iterative root finding algorithm such as Newton's method, and plugging them into Eq.~\eqref{pCCD_e_corr}. In order to facilitate the convergence of Newton's method, a good initial guess for each $t$ is necessary, which is informed here by many-body perturbation theory.

\section{Many-Body Perturbation Theory}\label{sec:manybody}

Many-body perturbation theory (MBPT) \cite{Paldus2006,shavittbartlett2009,morten_book} adds correlations between particles as perturbations to the Hartree-Fock wave function, the ground state of a collection of non-interacting particles subject to a mean-field potential that approximates their interaction. Analogously to single-body perturbation theory, it is possible to approximate the energy of a system by writing the Hamiltonian as the sum of two terms, one of which is readily diagonalizable. In this theory, we assume that the Hamiltonian $H$ can be written as the sum of an unperturbed Hamiltonian $H_0$ and an interacting Hamiltonian $H_I$, namely $H=H_0+H_I$, where $H_0\ket{\Phi_n}=E_n\ket{\Phi_n}$ is easily solvable.
Expanding the exact wave-function $\ket{\Psi_0}$ in terms of the unperturbed wave-function and its excited states and the defining the projection operators
$P=\ketbra{\Phi_0}$ and $Q=\sum_{n=1}^\infty\ketbra{\Phi_n}$, we can expand the correlation energy $\Delta E$ perturbatively in terms of the interaction $H_I$. A standard approach for this perturbative expansion is given by Rayleigh-Schr\"odinger (RS) perturbation theory, see for example Refs.~\cite{Paldus2006,shavittbartlett2009,morten_book}.
With RS many-body perturbation theory the correlation energy reads
\begin{align}
\label{e_sum}
\Delta E=\sum_{n=1}^\infty \Delta E^{(n)}\ ,
\end{align}
where, up to second-order in the interaction, we have
\begin{align}
\label{delta_e1}
\Delta E^{(1)}
=& \ 
\mel{\Phi_0}{H_I}{\Phi_0}\ ,
\\
\label{delta_e2}
\Delta E^{(2)}
=& \ 
\mel{\Phi_0}{H_I\frac{Q}{E_0-H_0}H_I}{\Phi_0}\ .
\end{align}
For the pairing model Hamiltonian (\ref{pairing_model_hamiltonian_original})
with the partitioning 
\begin{align}
\label{h0_hf}
H_0
=
\sum_{p\sigma}\epsilon_p a^\dagger_{p\sigma}a_{p\sigma}
\ ,
\end{align}
where
\begin{equation}
\epsilon_p = d_p-\sum_i\delta_{pi}g_{pp}\ ,
\end{equation}
we can rewrite the second order correlation energy contribution (\ref{delta_e2}) as
\begin{align}
\label{pairing_delta_e2}
\Delta E^{(2)}
&=
\frac{1}{2}\sum_{ia}\frac{g_{ai}g_{ia}}{d_i-d_a-g_{ii}}\ .
\end{align}
We can use this result to inform a good initial guess for the cluster parameters $t$ by comparing the pCCD correlation energy (\ref{pCCD_e_corr}) with the second order contribution to the correlation energy from MBPT, which allows us to identify
\begin{align}
\label{t_init}
{t_{ai}}^{(0)}
=
-\frac{1}{2}\frac{g_{ai}}{d_i-d_a-g_{ii}}\ .
\end{align}

\section{Results}
\label{sec:results}
\begin{figure}[t]
    \centering
    \includegraphics[width=0.48\textwidth]{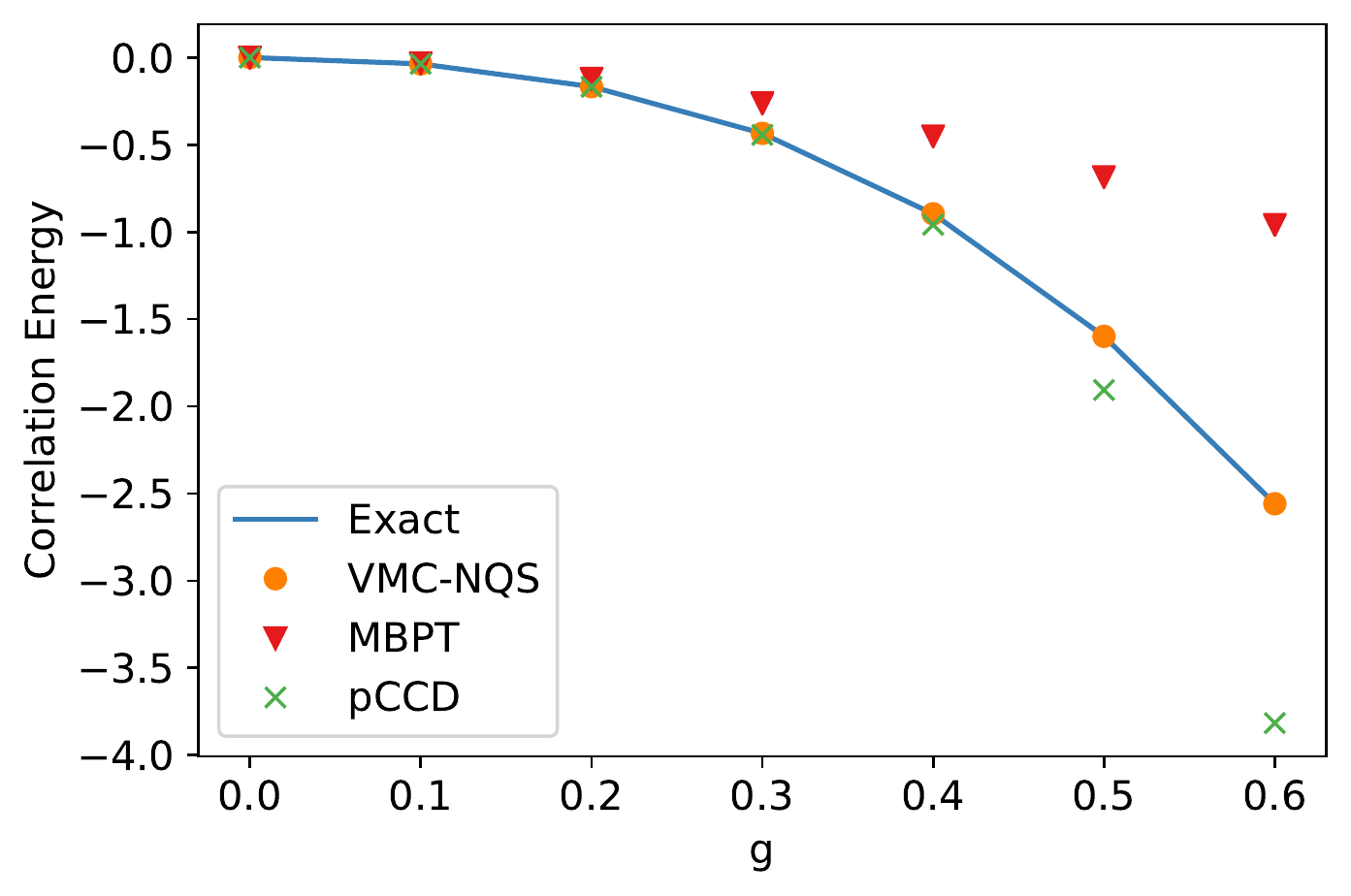}
    \caption{Correlation energy for the constant-coupling model for $P=10$ energy levels and $M=5$ pairs as a function of the interaction strength. VMC-NQS (solid orange circles), MBPT (solid red triangles), and pCCD (green crosses) calculations are compared with the exact answer (solid line).}
    \label{fig:corr_constant}
\end{figure}

The VMC-NQS values reported in this section were obtained by averaging over the energy estimates of 100 optimization steps at convergence. We initially present results for the Hamiltonian with constant pairing strength $g$ of Eq.~\eqref{pairing_model_hamiltonian_constant_g} considering $P=10$ energy levels occupied by $M=5$ pairs of fermions. The correlation energies as functions of the pairing strength computed within MBPT, pCCD, and VMC-NQS are displayed in Fig.~\ref{fig:corr_constant}. At small values of $g$, all many-body methods yield similar correlation energies. Already at $g \lesssim - 0.2 $, MBPT results start deviating from the exact-diagonalization ones, indicating the shortcomings of perturbation theory in addressing the large negative coupling constant region. A similar trend is observed by the pCCD calculations, albeit they remain closer to the exact ones down to $g\lesssim -0.4$. At deeper values of $g$, the pCCD energies become significantly lower than the exact ones, showing a clear violation of the variational principle. On the other hand, the correlation energies computed using VMC-NQS are in excellent agreement with the exact ones over the whole range of $g$ values considered and beyond, with discrepancies remaining always smaller than $10^{-4}$. Crucially, in contrast with MBPT and pCCD methods, VMC-NQS fulfils by construction the variational principle.  

\begin{figure}[t]
    \centering
    \includegraphics[width=0.48\textwidth]{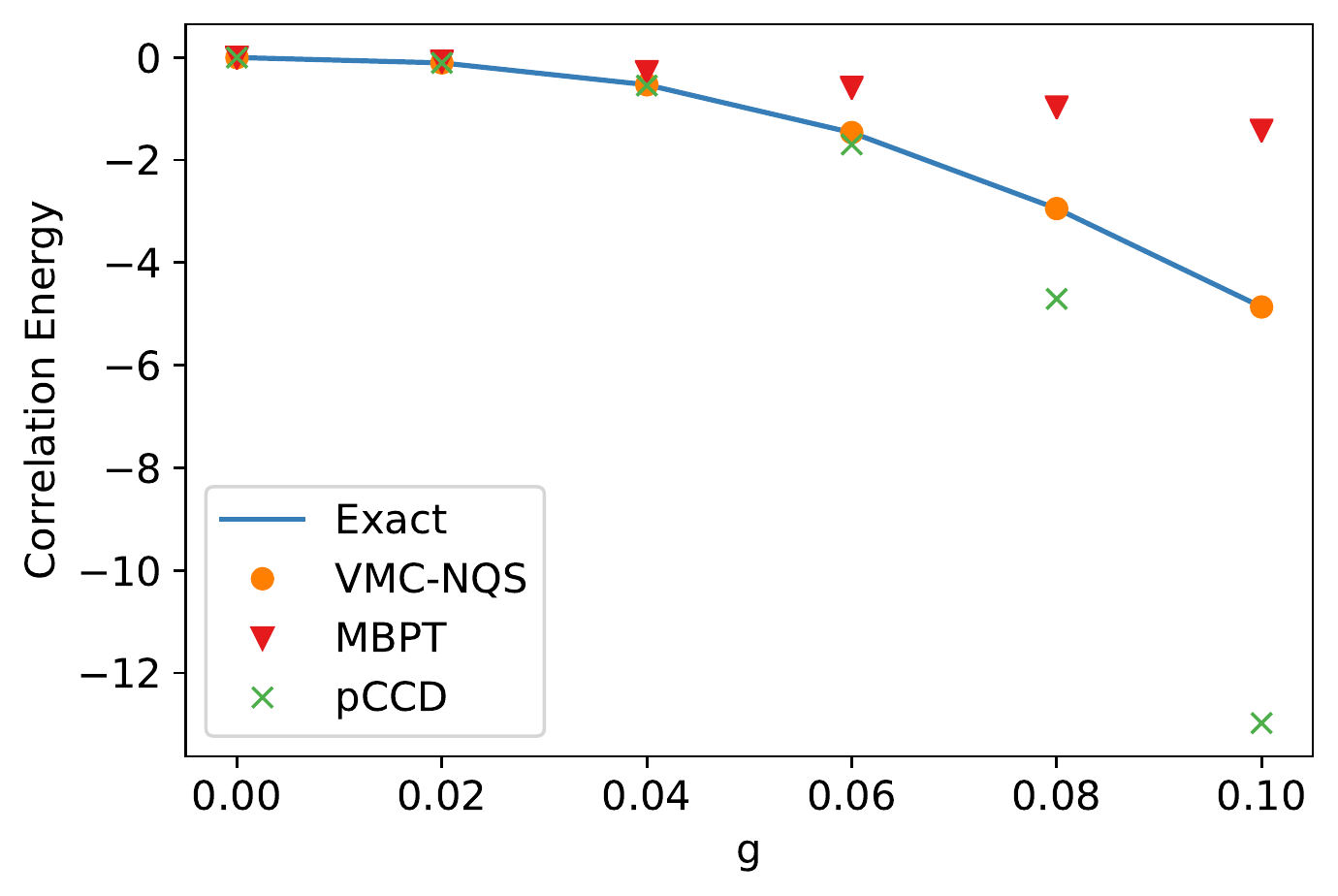}
    \caption{Correlation energy for the separable-coupling model for $P=10$ energy levels and $M=5$ pairs as a function of the interaction strength. VMC-NQS (solid orange circles), MBPT (solid red triangles), and pCCD (green crosses) calculations are compared with the exact answer (solid line).}
    \label{fig:corr_separable}
\end{figure}

We then extend our comparative analysis considering a larger model space, with up to $P=40$ energy levels and $M=20$ pairs of fermions. In Table~\ref{tab:g_constant_large}, we compare the correlation energies obtained within VMC-NQS, pCCD, and the highly-accurate iterative approach for solving the Richardson equations developed in Ref.~\cite{Guan:2022CoPhC}. To make contact with the results of the latter reference, the single-particle energies are re-scaled as $d_p = p / 10$ and the pairing strength is taken to be $g=-0.05$. Similarly to the smaller model space, the correlation energies computed within VMC-NQS are almost always in perfect agreement with the ones obtained from the Iterative method. The only exception is the case with $P=40$ and $M=20$, for which the VMC-NQS yields $-1.785$, while the iterative method provides $-2.111$. In this regard, we successfully executed the MATHEMATICA program attached to Ref.~\cite{Guan:2022CoPhC} for all combinations of $P$ and $M$, except for $P=40$ and $M=20$, where we could not obtain converged results. On the other hand, pCCD struggles already for $P=10$ and $M=5$, where it yields $\sim 20\%$ overbinding. Increasing the number of single-particle states and pairs, pCCD becomes more and more unreliable, with large departures from both the exact and VMC-NQS values. This behavior is due to the the pairing interaction terms growing faster with the number of states than the mean-field one, making the exact ground-state very different from the Hartree-Fock solution.
\begin{table}[b]
\begin{center}
\vspace{0.5cm}
\begin{tabular}{ c | c | c c c c }
\hline
\hline
& $P$ & $M=5$  & $M=10$ & $M=15$ & $M=20$ \\
\hline
VMC-NQS & \multirow{3}{*}{ 10 }  & -0.160 & 0.000 & - & - \\
pCCD &  & -0.193 & 0.000 & - &  -\\
Iterative &  & -0.160  & 0.000 & - & - \\
\hline
VMC-NQS & \multirow{3}{*}{ 20 }  & -0.394 & -0.515 & -0.394 & 0.000 \\
pCCD &   & -0.987 & -1.743 & -1.110 &  0.000 \\
Iterative &   & -0.394 & -0.515 & -0.394 &  0.000 \\
\hline
VMC-NQS & \multirow{3}{*}{ 30 }  & -0.606 & -0.946 & -1.057 & -0.946 \\
pCCD &   & -1.750 & 3.500 & -6.467 &  -5.697\\
Iterative &   & -0.606 & -0.946 & -1.057 & -0.946\\
\hline
VMC-NQS & \multirow{3}{*}{ 40 }  & -0.809 & -1.356 & -1.678 & -1.785 \\
pCCD &   & -4.231 & 124.923 & 7.598 &  -44.346 \\
Iterative &   & -0.809 & -1.356 & -1.678 & -2.110\\
\hline
\end{tabular}
\caption{Correlation energies obtained with the VMC-NQS method compared with pCCD and the iterative approach of Ref.~\cite{Guan:2022CoPhC} for the constant-coupling Hamiltonian of Eq.~\eqref{pairing_model_hamiltonian_constant_g} with $g = 0.05$ and $d_p=p/10$.}
\label{tab:g_constant_large}
\end{center}
\end{table}
We now turn to the separable-pairing Hamiltonian of Eq.~\eqref{pairing_model_hamiltonian_separable_g}. In Fig.~\ref{fig:corr_separable}, we display the correlation energy for this Hamiltonian for $P=10$ energy levels and $M=5$ pairs as a function of the interaction strength. Consistent with Ref.~\cite{Dukelsky:2011nb}, we use the cutoff $\alpha=10$. The exact-diagonalization and VMC-NQS results overlaps nicely over the entire range of pairing strength that we consider, in some cases matching up to 8 significant digits, corroborating once more the accuracy of NQS in representing complicated many-body wave functions. Similarly to the results displayed in Fig.~\ref{fig:corr_constant} for the constant-pairing case, MBPT yields accurate correlation energies only for very small values $g$, while for larger values of the pairing strength large deviations are observed. The energies computed within pCCD start deviating from the exact ones for $g\gtrsim 0.06$, exhibiting abrupt violations of the variational principle. Consistent with the constant-pairing energies listed in Table~\ref{tab:g_constant_large}, pCCD becomes less accurate with increasing system size, while VMC-NQS does not seem to suffer similar limitations. In addition, we verified that the nearly perfect agreement between VMC-NQS and exact-diagonalization  energies persist at least up until $g=0.6$, confirming once again the robustness of the method in the non-perturbative regime. 

\section{Conclusions}
\label{sec:conclusions}
We have introduced a variational Monte Carlo method based on neural-network quantum states that solves the nuclear many-body problem in the occupation-number formalism. A tailored version of the stochastic-reconfiguration algorithm, with a regularization term inspired by RMSprop, is utilized to train the neural networks and minimize the Hamiltonian expectation value. 

As a specific application of this method, which exhibits a polynomial scaling with the number of single-particle levels, we considered two classes of exactly solvable pairing models both with constant and separable pairing interaction strength. In addition to exact-diagonalization techniques, which are limited to relatively small model spaces, we benchmark the VMC-NQS approach against virtually-exact methods that solve the Gaudin-Richardson equations in an iterative fashion. We also gauge the accuracy of many-body perturbation theory and pair coupled cluster doubles theory, both routinely used to solve the nuclear many-body problem.  

We find the VMC-NQS results to be in excellent agreement with the exact solution, independent of the magnitude of the pairing strength. As expected, the correlation energies obtained from MBPT start deviating from the exact values when the pairing strength becomes large. A similar trend is observed by pCCD calculations, although they remain closer to the exact solutions for a broader range of pairing strengths values. Most notably, VMC-NQS is guaranteed to respect the variational principle, which is in general violated by both MBPT and pCCD. 

It has to be mentioned that VMC-NQS is not limited to constant or separable pairing interactions. As an immediate application of this method, we will consider the state dependent, non separable, Gogny interaction in the Hartree-Fock basis and compute the ground-state energies of nuclei going beyond the mean-field approximation. In this regard, it would be interesting to validate the conclusions of Ref.~\cite{Dukelsky:2011nb}, which were obtained by fitting an exactly solvable hyperbolic model to the gaps and pairing tensors of Gogny Hartree-Fock-Bogoliubov calculations.

The applicability of VMC-NQS extends to pairing Hamiltonians relevant to model interacting systems in condensed matter. An example is the characterization of number-conserving topological superconductors, or superfluids, and the fate of Majorana zero-modes beyond mean-field~\cite{Ortiz:2014PhRvL}. While the Richardson-Gaudin-Kitaev model is integrable for periodic and antiperiodic boundary conditions, no exact solutions are known for open boundaries. Since the interactions are long-range, DMRG approaches struggle to converge~\cite{Iemini:2015PhRvL}, while deep neural quantum states should be able to provide more accurate solutions.

Although interesting in its own right, solving the pairing model is only the first application of NQS to the nuclear many body problem in the occupation number formalism. As a next step, we will consider more realistic shell-model Hamiltonians~\cite{hko95}. We plan on carrying out benchmark calculations with exact-diagonalization methods and coupled cluster theory in  pf -shell nuclei. The long-range goal is to provide accurate solutions to Hamiltonians that are systematically derived within chiral effective field theory, extending the work of Ref.~\cite{Tichai:2022bxr} that was limited to a matrix product state ansatz.

\section*{Acknowledgements}
We thank Y. Alexeev, I. Martin, and G. Ortiz for stimulating discussions. The present research is supported by the U.S. Department of Energy, Office of Science, Office of Nuclear Physics, under contracts DE-AC02-06CH11357 (A.L), by the 2020 DOE Early Career Award number ANL PRJ1008597 (A.L.), the NUCLEI SciDAC program (A.L.), and Argonne LDRD awards (A. L.). 
B.H. and M.H-J. are supported by the U.S. Department of Energy,
Office of Science, office of Nuclear Physics under grant
No. DE-SC0021152 and U.S. National Science Foundation Grants
No. PHY-1404159 and PHY-2013047.

\bibliography{biblio}

\begin{thebibliography}{56}%
\makeatletter
\providecommand \@ifxundefined [1]{%
 \@ifx{#1\undefined}
}%
\providecommand \@ifnum [1]{%
 \ifnum #1\expandafter \@firstoftwo
 \else \expandafter \@secondoftwo
 \fi
}%
\providecommand \@ifx [1]{%
 \ifx #1\expandafter \@firstoftwo
 \else \expandafter \@secondoftwo
 \fi
}%
\providecommand \natexlab [1]{#1}%
\providecommand \enquote  [1]{``#1''}%
\providecommand \bibnamefont  [1]{#1}%
\providecommand \bibfnamefont [1]{#1}%
\providecommand \citenamefont [1]{#1}%
\providecommand \href@noop [0]{\@secondoftwo}%
\providecommand \href [0]{\begingroup \@sanitize@url \@href}%
\providecommand \@href[1]{\@@startlink{#1}\@@href}%
\providecommand \@@href[1]{\endgroup#1\@@endlink}%
\providecommand \@sanitize@url [0]{\catcode `\\12\catcode `\$12\catcode
  `\&12\catcode `\#12\catcode `\^12\catcode `\_12\catcode `\%12\relax}%
\providecommand \@@startlink[1]{}%
\providecommand \@@endlink[0]{}%
\providecommand \url  [0]{\begingroup\@sanitize@url \@url }%
\providecommand \@url [1]{\endgroup\@href {#1}{\urlprefix }}%
\providecommand \urlprefix  [0]{URL }%
\providecommand \Eprint [0]{\href }%
\providecommand \doibase [0]{http://dx.doi.org/}%
\providecommand \selectlanguage [0]{\@gobble}%
\providecommand \bibinfo  [0]{\@secondoftwo}%
\providecommand \bibfield  [0]{\@secondoftwo}%
\providecommand \translation [1]{[#1]}%
\providecommand \BibitemOpen [0]{}%
\providecommand \bibitemStop [0]{}%
\providecommand \bibitemNoStop [0]{.\EOS\space}%
\providecommand \EOS [0]{\spacefactor3000\relax}%
\providecommand \BibitemShut  [1]{\csname bibitem#1\endcsname}%
\let\auto@bib@innerbib\@empty
%</preamble>
\bibitem [{\citenamefont {Hagen}\ \emph {et~al.}(2014)\citenamefont {Hagen},
  \citenamefont {Papenbrock}, \citenamefont {Hjorth-Jensen},\ and\
  \citenamefont {Dean}}]{Hagen:2013nca}%
  \BibitemOpen
  \bibfield  {author} {\bibinfo {author} {\bibfnamefont {G.}~\bibnamefont
  {Hagen}}, \bibinfo {author} {\bibfnamefont {T.}~\bibnamefont {Papenbrock}},
  \bibinfo {author} {\bibfnamefont {M.}~\bibnamefont {Hjorth-Jensen}}, \ and\
  \bibinfo {author} {\bibfnamefont {D.~J.}\ \bibnamefont {Dean}},\ }\href
  {\doibase 10.1088/0034-4885/77/9/096302} {\bibfield  {journal} {\bibinfo
  {journal} {Rept. Prog. Phys.}\ }\textbf {\bibinfo {volume} {77}},\ \bibinfo
  {pages} {096302} (\bibinfo {year} {2014})},\ \Eprint
  {http://arxiv.org/abs/1312.7872} {arXiv:1312.7872 [nucl-th]} \BibitemShut
  {NoStop}%
\bibitem [{\citenamefont {Hergert}\ \emph {et~al.}(2016)\citenamefont
  {Hergert}, \citenamefont {Bogner}, \citenamefont {Morris}, \citenamefont
  {Schwenk},\ and\ \citenamefont {Tsukiyama}}]{Hergert:2015awm}%
  \BibitemOpen
  \bibfield  {author} {\bibinfo {author} {\bibfnamefont {H.}~\bibnamefont
  {Hergert}}, \bibinfo {author} {\bibfnamefont {S.~K.}\ \bibnamefont {Bogner}},
  \bibinfo {author} {\bibfnamefont {T.~D.}\ \bibnamefont {Morris}}, \bibinfo
  {author} {\bibfnamefont {A.}~\bibnamefont {Schwenk}}, \ and\ \bibinfo
  {author} {\bibfnamefont {K.}~\bibnamefont {Tsukiyama}},\ }\href {\doibase
  10.1016/j.physrep.2015.12.007} {\bibfield  {journal} {\bibinfo  {journal}
  {Phys. Rept.}\ }\textbf {\bibinfo {volume} {621}},\ \bibinfo {pages} {165}
  (\bibinfo {year} {2016})},\ \Eprint {http://arxiv.org/abs/1512.06956}
  {arXiv:1512.06956 [nucl-th]} \BibitemShut {NoStop}%
\bibitem [{\citenamefont {Som\`a}(2020)}]{Soma:2020xhv}%
  \BibitemOpen
  \bibfield  {author} {\bibinfo {author} {\bibfnamefont {V.}~\bibnamefont
  {Som\`a}},\ }\href {\doibase 10.3389/fphy.2020.00340} {\bibfield  {journal}
  {\bibinfo  {journal} {Front. in Phys.}\ }\textbf {\bibinfo {volume} {8}},\
  \bibinfo {pages} {340} (\bibinfo {year} {2020})},\ \Eprint
  {http://arxiv.org/abs/2003.11321} {arXiv:2003.11321 [nucl-th]} \BibitemShut
  {NoStop}%
\bibitem [{\citenamefont {Lee}(2020)}]{Lee:2020meg}%
  \BibitemOpen
  \bibfield  {author} {\bibinfo {author} {\bibfnamefont {D.}~\bibnamefont
  {Lee}},\ }\href {\doibase 10.3389/fphy.2020.00174} {\bibfield  {journal}
  {\bibinfo  {journal} {Front. in Phys.}\ }\textbf {\bibinfo {volume} {8}},\
  \bibinfo {pages} {174} (\bibinfo {year} {2020})}\BibitemShut {NoStop}%
\bibitem [{\citenamefont {Carlson}\ \emph {et~al.}(2015)\citenamefont
  {Carlson}, \citenamefont {Gandolfi}, \citenamefont {Pederiva}, \citenamefont
  {Pieper}, \citenamefont {Schiavilla}, \citenamefont {Schmidt},\ and\
  \citenamefont {Wiringa}}]{Carlson:2014vla}%
  \BibitemOpen
  \bibfield  {author} {\bibinfo {author} {\bibfnamefont {J.}~\bibnamefont
  {Carlson}}, \bibinfo {author} {\bibfnamefont {S.}~\bibnamefont {Gandolfi}},
  \bibinfo {author} {\bibfnamefont {F.}~\bibnamefont {Pederiva}}, \bibinfo
  {author} {\bibfnamefont {S.~C.}\ \bibnamefont {Pieper}}, \bibinfo {author}
  {\bibfnamefont {R.}~\bibnamefont {Schiavilla}}, \bibinfo {author}
  {\bibfnamefont {K.~E.}\ \bibnamefont {Schmidt}}, \ and\ \bibinfo {author}
  {\bibfnamefont {R.~B.}\ \bibnamefont {Wiringa}},\ }\href {\doibase
  10.1103/RevModPhys.87.1067} {\bibfield  {journal} {\bibinfo  {journal} {Rev.
  Mod. Phys.}\ }\textbf {\bibinfo {volume} {87}},\ \bibinfo {pages} {1067}
  (\bibinfo {year} {2015})},\ \Eprint {http://arxiv.org/abs/1412.3081}
  {arXiv:1412.3081 [nucl-th]} \BibitemShut {NoStop}%
\bibitem [{\citenamefont {Gysbers}\ \emph {et~al.}(2019)\citenamefont {Gysbers}
  \emph {et~al.}}]{Gysbers:2019uyb}%
  \BibitemOpen
  \bibfield  {author} {\bibinfo {author} {\bibfnamefont {P.}~\bibnamefont
  {Gysbers}} \emph {et~al.},\ }\href {\doibase 10.1038/s41567-019-0450-7}
  {\bibfield  {journal} {\bibinfo  {journal} {Nature Phys.}\ }\textbf {\bibinfo
  {volume} {15}},\ \bibinfo {pages} {428} (\bibinfo {year} {2019})},\ \Eprint
  {http://arxiv.org/abs/1903.00047} {arXiv:1903.00047 [nucl-th]} \BibitemShut
  {NoStop}%
\bibitem [{\citenamefont {Malbrunot-Ettenauer}\ \emph
  {et~al.}(2022)\citenamefont {Malbrunot-Ettenauer} \emph
  {et~al.}}]{Malbrunot-Ettenauer:2021fnr}%
  \BibitemOpen
  \bibfield  {author} {\bibinfo {author} {\bibfnamefont {S.}~\bibnamefont
  {Malbrunot-Ettenauer}} \emph {et~al.},\ }\href {\doibase
  10.1103/PhysRevLett.128.022502} {\bibfield  {journal} {\bibinfo  {journal}
  {Phys. Rev. Lett.}\ }\textbf {\bibinfo {volume} {128}},\ \bibinfo {pages}
  {022502} (\bibinfo {year} {2022})},\ \Eprint
  {http://arxiv.org/abs/2112.03382} {arXiv:2112.03382 [nucl-ex]} \BibitemShut
  {NoStop}%
\bibitem [{\citenamefont {Hu}\ \emph {et~al.}(2021)\citenamefont {Hu} \emph
  {et~al.}}]{Hu:2021trw}%
  \BibitemOpen
  \bibfield  {author} {\bibinfo {author} {\bibfnamefont {B.}~\bibnamefont {Hu}}
  \emph {et~al.},\ }\href {\doibase 10.1038/s41567-022-01715-8} {\  (\bibinfo
  {year} {2021}),\ 10.1038/s41567-022-01715-8},\ \Eprint
  {http://arxiv.org/abs/2112.01125} {arXiv:2112.01125 [nucl-th]} \BibitemShut
  {NoStop}%
\bibitem [{\citenamefont {Carleo}\ and\ \citenamefont
  {Troyer}(2017)}]{Carleo:2017}%
  \BibitemOpen
  \bibfield  {author} {\bibinfo {author} {\bibfnamefont {G.}~\bibnamefont
  {Carleo}}\ and\ \bibinfo {author} {\bibfnamefont {M.}~\bibnamefont
  {Troyer}},\ }\href {\doibase 10.1126/science.aag2302} {\bibfield  {journal}
  {\bibinfo  {journal} {Science}\ }\textbf {\bibinfo {volume} {355}},\ \bibinfo
  {pages} {602} (\bibinfo {year} {2017})}\BibitemShut {NoStop}%
\bibitem [{\citenamefont {Choo}\ \emph {et~al.}(2020)\citenamefont {Choo},
  \citenamefont {Mezzacapo},\ and\ \citenamefont {Carleo}}]{Choo:2019}%
  \BibitemOpen
  \bibfield  {author} {\bibinfo {author} {\bibfnamefont {K.}~\bibnamefont
  {Choo}}, \bibinfo {author} {\bibfnamefont {A.}~\bibnamefont {Mezzacapo}}, \
  and\ \bibinfo {author} {\bibfnamefont {G.}~\bibnamefont {Carleo}},\ }\href
  {\doibase 10.1038/s41467-020-15724-9} {\bibfield  {journal} {\bibinfo
  {journal} {Nature Communications}\ }\textbf {\bibinfo {volume} {11}},\
  \bibinfo {pages} {2368} (\bibinfo {year} {2020})}\BibitemShut {NoStop}%
\bibitem [{\citenamefont {{Hermann}}\ \emph {et~al.}(2020)\citenamefont
  {{Hermann}}, \citenamefont {{Sch{\"a}tzle}},\ and\ \citenamefont
  {{No{\'e}}}}]{Hermann:2019}%
  \BibitemOpen
  \bibfield  {author} {\bibinfo {author} {\bibfnamefont {J.}~\bibnamefont
  {{Hermann}}}, \bibinfo {author} {\bibfnamefont {Z.}~\bibnamefont
  {{Sch{\"a}tzle}}}, \ and\ \bibinfo {author} {\bibfnamefont {F.}~\bibnamefont
  {{No{\'e}}}},\ }\href {\doibase 10.1038/s41557-020-0544-y} {\bibfield
  {journal} {\bibinfo  {journal} {Nature Chemistry}\ }\textbf {\bibinfo
  {volume} {12}},\ \bibinfo {pages} {891} (\bibinfo {year} {2020})}\BibitemShut
  {NoStop}%
\bibitem [{\citenamefont {{Pfau}}\ \emph {et~al.}(2020)\citenamefont {{Pfau}},
  \citenamefont {{Spencer}}, \citenamefont {{Matthews}},\ and\ \citenamefont
  {{Foulkes}}}]{Pfau:2019}%
  \BibitemOpen
  \bibfield  {author} {\bibinfo {author} {\bibfnamefont {D.}~\bibnamefont
  {{Pfau}}}, \bibinfo {author} {\bibfnamefont {J.~S.}\ \bibnamefont
  {{Spencer}}}, \bibinfo {author} {\bibfnamefont {A.~G.~D.~G.}\ \bibnamefont
  {{Matthews}}}, \ and\ \bibinfo {author} {\bibfnamefont {W.~M.~C.}\
  \bibnamefont {{Foulkes}}},\ }\href {\doibase
  10.1103/PhysRevResearch.2.033429} {\bibfield  {journal} {\bibinfo  {journal}
  {Physical Review Research}\ }\textbf {\bibinfo {volume} {2}},\ \bibinfo {eid}
  {033429} (\bibinfo {year} {2020})},\ \Eprint
  {http://arxiv.org/abs/1909.02487} {arXiv:1909.02487 [physics.chem-ph]}
  \BibitemShut {NoStop}%
\bibitem [{\citenamefont {Pescia}\ \emph {et~al.}(2022)\citenamefont {Pescia},
  \citenamefont {Han}, \citenamefont {Lovato}, \citenamefont {Lu},\ and\
  \citenamefont {Carleo}}]{Pescia:2021kxb}%
  \BibitemOpen
  \bibfield  {author} {\bibinfo {author} {\bibfnamefont {G.}~\bibnamefont
  {Pescia}}, \bibinfo {author} {\bibfnamefont {J.}~\bibnamefont {Han}},
  \bibinfo {author} {\bibfnamefont {A.}~\bibnamefont {Lovato}}, \bibinfo
  {author} {\bibfnamefont {J.}~\bibnamefont {Lu}}, \ and\ \bibinfo {author}
  {\bibfnamefont {G.}~\bibnamefont {Carleo}},\ }\href {\doibase
  10.1103/PhysRevResearch.4.023138} {\bibfield  {journal} {\bibinfo  {journal}
  {Phys. Rev. Res.}\ }\textbf {\bibinfo {volume} {4}},\ \bibinfo {pages}
  {023138} (\bibinfo {year} {2022})},\ \Eprint
  {http://arxiv.org/abs/2112.11957} {arXiv:2112.11957 [quant-ph]} \BibitemShut
  {NoStop}%
\bibitem [{\citenamefont {Wilson}\ \emph {et~al.}(2022)\citenamefont {Wilson},
  \citenamefont {Moroni}, \citenamefont {Holzmann}, \citenamefont {Gao},
  \citenamefont {Wudarski}, \citenamefont {Vegge},\ and\ \citenamefont
  {Bhowmik}}]{Wilson:2022meh}%
  \BibitemOpen
  \bibfield  {author} {\bibinfo {author} {\bibfnamefont {M.}~\bibnamefont
  {Wilson}}, \bibinfo {author} {\bibfnamefont {S.}~\bibnamefont {Moroni}},
  \bibinfo {author} {\bibfnamefont {M.}~\bibnamefont {Holzmann}}, \bibinfo
  {author} {\bibfnamefont {N.}~\bibnamefont {Gao}}, \bibinfo {author}
  {\bibfnamefont {F.}~\bibnamefont {Wudarski}}, \bibinfo {author}
  {\bibfnamefont {T.}~\bibnamefont {Vegge}}, \ and\ \bibinfo {author}
  {\bibfnamefont {A.}~\bibnamefont {Bhowmik}},\ }\href@noop {} {\  (\bibinfo
  {year} {2022})},\ \Eprint {http://arxiv.org/abs/2202.04622} {arXiv:2202.04622
  [physics.chem-ph]} \BibitemShut {NoStop}%
\bibitem [{\citenamefont {Stokes}\ \emph {et~al.}(2020)\citenamefont {Stokes},
  \citenamefont {Moreno}, \citenamefont {Pnevmatikakis},\ and\ \citenamefont
  {Carleo}}]{Stokes:2020ihf}%
  \BibitemOpen
  \bibfield  {author} {\bibinfo {author} {\bibfnamefont {J.}~\bibnamefont
  {Stokes}}, \bibinfo {author} {\bibfnamefont {J.~R.}\ \bibnamefont {Moreno}},
  \bibinfo {author} {\bibfnamefont {E.~A.}\ \bibnamefont {Pnevmatikakis}}, \
  and\ \bibinfo {author} {\bibfnamefont {G.}~\bibnamefont {Carleo}},\ }\href
  {\doibase 10.1103/PhysRevB.102.205122} {\bibfield  {journal} {\bibinfo
  {journal} {Phys. Rev. B}\ }\textbf {\bibinfo {volume} {102}},\ \bibinfo
  {pages} {205122} (\bibinfo {year} {2020})},\ \Eprint
  {http://arxiv.org/abs/2008.00118} {arXiv:2008.00118 [cond-mat.str-el]}
  \BibitemShut {NoStop}%
\bibitem [{\citenamefont {Moreno}\ \emph {et~al.}(2022)\citenamefont {Moreno},
  \citenamefont {Carleo}, \citenamefont {Georges},\ and\ \citenamefont
  {Stokes}}]{Moreno:2021jas}%
  \BibitemOpen
  \bibfield  {author} {\bibinfo {author} {\bibfnamefont {J.~R.}\ \bibnamefont
  {Moreno}}, \bibinfo {author} {\bibfnamefont {G.}~\bibnamefont {Carleo}},
  \bibinfo {author} {\bibfnamefont {A.}~\bibnamefont {Georges}}, \ and\
  \bibinfo {author} {\bibfnamefont {J.}~\bibnamefont {Stokes}},\ }\href
  {\doibase 10.1073/pnas.2122059119} {\bibfield  {journal} {\bibinfo  {journal}
  {Proc. Nat. Acad. Sci.}\ }\textbf {\bibinfo {volume} {119}},\ \bibinfo
  {pages} {e2122059119} (\bibinfo {year} {2022})},\ \Eprint
  {http://arxiv.org/abs/2111.10420} {arXiv:2111.10420 [cond-mat.str-el]}
  \BibitemShut {NoStop}%
\bibitem [{\citenamefont {{Zhao}}\ \emph {et~al.}(2022)\citenamefont {{Zhao}},
  \citenamefont {{Stokes}},\ and\ \citenamefont {{Veerapaneni}}}]{Zhao:2022}%
  \BibitemOpen
  \bibfield  {author} {\bibinfo {author} {\bibfnamefont {T.}~\bibnamefont
  {{Zhao}}}, \bibinfo {author} {\bibfnamefont {J.}~\bibnamefont {{Stokes}}}, \
  and\ \bibinfo {author} {\bibfnamefont {S.}~\bibnamefont {{Veerapaneni}}},\
  }\href@noop {} {\bibfield  {journal} {\bibinfo  {journal} {arXiv e-prints}\
  ,\ \bibinfo {eid} {arXiv:2208.05637}} (\bibinfo {year} {2022})},\ \Eprint
  {http://arxiv.org/abs/2208.05637} {arXiv:2208.05637 [physics.chem-ph]}
  \BibitemShut {NoStop}%
\bibitem [{\citenamefont {Keeble}\ and\ \citenamefont
  {Rios}(2020)}]{Keeble:2019bkv}%
  \BibitemOpen
  \bibfield  {author} {\bibinfo {author} {\bibfnamefont {J.~W.~T.}\
  \bibnamefont {Keeble}}\ and\ \bibinfo {author} {\bibfnamefont
  {A.}~\bibnamefont {Rios}},\ }\href {\doibase 10.1016/j.physletb.2020.135743}
  {\bibfield  {journal} {\bibinfo  {journal} {Phys. Lett. B}\ }\textbf
  {\bibinfo {volume} {809}},\ \bibinfo {pages} {135743} (\bibinfo {year}
  {2020})},\ \Eprint {http://arxiv.org/abs/1911.13092} {arXiv:1911.13092
  [nucl-th]} \BibitemShut {NoStop}%
\bibitem [{\citenamefont {Sarmiento}\ \emph {et~al.}(2022)\citenamefont
  {Sarmiento}, \citenamefont {Keeble},\ and\ \citenamefont
  {Rios}}]{Sarmiento:2022bxn}%
  \BibitemOpen
  \bibfield  {author} {\bibinfo {author} {\bibfnamefont {J.~R.}\ \bibnamefont
  {Sarmiento}}, \bibinfo {author} {\bibfnamefont {J.~W.~T.}\ \bibnamefont
  {Keeble}}, \ and\ \bibinfo {author} {\bibfnamefont {A.}~\bibnamefont
  {Rios}},\ }\href@noop {} {\  (\bibinfo {year} {2022})},\ \Eprint
  {http://arxiv.org/abs/2205.12795} {arXiv:2205.12795 [nucl-th]} \BibitemShut
  {NoStop}%
\bibitem [{\citenamefont {Adams}\ \emph {et~al.}(2021)\citenamefont {Adams},
  \citenamefont {Carleo}, \citenamefont {Lovato},\ and\ \citenamefont
  {Rocco}}]{Adams:2020aax}%
  \BibitemOpen
  \bibfield  {author} {\bibinfo {author} {\bibfnamefont {C.}~\bibnamefont
  {Adams}}, \bibinfo {author} {\bibfnamefont {G.}~\bibnamefont {Carleo}},
  \bibinfo {author} {\bibfnamefont {A.}~\bibnamefont {Lovato}}, \ and\ \bibinfo
  {author} {\bibfnamefont {N.}~\bibnamefont {Rocco}},\ }\href {\doibase
  10.1103/PhysRevLett.127.022502} {\bibfield  {journal} {\bibinfo  {journal}
  {Phys. Rev. Lett.}\ }\textbf {\bibinfo {volume} {127}},\ \bibinfo {pages}
  {022502} (\bibinfo {year} {2021})},\ \Eprint
  {http://arxiv.org/abs/2007.14282} {arXiv:2007.14282 [nucl-th]} \BibitemShut
  {NoStop}%
\bibitem [{\citenamefont {Gnech}\ \emph {et~al.}(2022)\citenamefont {Gnech},
  \citenamefont {Adams}, \citenamefont {Brawand}, \citenamefont {Carleo},
  \citenamefont {Lovato},\ and\ \citenamefont {Rocco}}]{Gnech:2021wfn}%
  \BibitemOpen
  \bibfield  {author} {\bibinfo {author} {\bibfnamefont {A.}~\bibnamefont
  {Gnech}}, \bibinfo {author} {\bibfnamefont {C.}~\bibnamefont {Adams}},
  \bibinfo {author} {\bibfnamefont {N.}~\bibnamefont {Brawand}}, \bibinfo
  {author} {\bibfnamefont {G.}~\bibnamefont {Carleo}}, \bibinfo {author}
  {\bibfnamefont {A.}~\bibnamefont {Lovato}}, \ and\ \bibinfo {author}
  {\bibfnamefont {N.}~\bibnamefont {Rocco}},\ }\href {\doibase
  10.1007/s00601-021-01706-0} {\bibfield  {journal} {\bibinfo  {journal} {Few
  Body Syst.}\ }\textbf {\bibinfo {volume} {63}},\ \bibinfo {pages} {7}
  (\bibinfo {year} {2022})},\ \Eprint {http://arxiv.org/abs/2108.06836}
  {arXiv:2108.06836 [nucl-th]} \BibitemShut {NoStop}%
\bibitem [{\citenamefont {Lovato}\ \emph {et~al.}(2022)\citenamefont {Lovato},
  \citenamefont {Adams}, \citenamefont {Carleo},\ and\ \citenamefont
  {Rocco}}]{Lovato:2022tjh}%
  \BibitemOpen
  \bibfield  {author} {\bibinfo {author} {\bibfnamefont {A.}~\bibnamefont
  {Lovato}}, \bibinfo {author} {\bibfnamefont {C.}~\bibnamefont {Adams}},
  \bibinfo {author} {\bibfnamefont {G.}~\bibnamefont {Carleo}}, \ and\ \bibinfo
  {author} {\bibfnamefont {N.}~\bibnamefont {Rocco}},\ }\href@noop {} {\
  (\bibinfo {year} {2022})},\ \Eprint {http://arxiv.org/abs/2206.10021}
  {arXiv:2206.10021 [nucl-th]} \BibitemShut {NoStop}%
\bibitem [{\citenamefont {Bohr}\ \emph {et~al.}(1958)\citenamefont {Bohr},
  \citenamefont {Mottelson},\ and\ \citenamefont {Pines}}]{Bohr:1958zz}%
  \BibitemOpen
  \bibfield  {author} {\bibinfo {author} {\bibfnamefont {A.}~\bibnamefont
  {Bohr}}, \bibinfo {author} {\bibfnamefont {B.~R.}\ \bibnamefont {Mottelson}},
  \ and\ \bibinfo {author} {\bibfnamefont {D.}~\bibnamefont {Pines}},\ }\href
  {\doibase 10.1103/PhysRev.110.936} {\bibfield  {journal} {\bibinfo  {journal}
  {Phys. Rev.}\ }\textbf {\bibinfo {volume} {110}},\ \bibinfo {pages} {936}
  (\bibinfo {year} {1958})}\BibitemShut {NoStop}%
\bibitem [{\citenamefont {Dean}\ and\ \citenamefont
  {Hjorth-Jensen}(2003)}]{Dean:2002zx}%
  \BibitemOpen
  \bibfield  {author} {\bibinfo {author} {\bibfnamefont {D.~J.}\ \bibnamefont
  {Dean}}\ and\ \bibinfo {author} {\bibfnamefont {M.}~\bibnamefont
  {Hjorth-Jensen}},\ }\href {\doibase 10.1103/RevModPhys.75.607} {\bibfield
  {journal} {\bibinfo  {journal} {Rev. Mod. Phys.}\ }\textbf {\bibinfo {volume}
  {75}},\ \bibinfo {pages} {607} (\bibinfo {year} {2003})},\ \Eprint
  {http://arxiv.org/abs/nucl-th/0210033} {arXiv:nucl-th/0210033} \BibitemShut
  {NoStop}%
\bibitem [{\citenamefont {Moller}\ and\ \citenamefont
  {Nix}(1992)}]{Moller:1992zz}%
  \BibitemOpen
  \bibfield  {author} {\bibinfo {author} {\bibfnamefont {P.}~\bibnamefont
  {Moller}}\ and\ \bibinfo {author} {\bibfnamefont {J.~R.}\ \bibnamefont
  {Nix}},\ }\href {\doibase 10.1016/0375-9474(92)90244-E} {\bibfield  {journal}
  {\bibinfo  {journal} {Nucl. Phys. A}\ }\textbf {\bibinfo {volume} {536}},\
  \bibinfo {pages} {20} (\bibinfo {year} {1992})}\BibitemShut {NoStop}%
\bibitem [{\citenamefont {Molique}\ and\ \citenamefont
  {Dudek}(1997)}]{Molique:1997zz}%
  \BibitemOpen
  \bibfield  {author} {\bibinfo {author} {\bibfnamefont {H.}~\bibnamefont
  {Molique}}\ and\ \bibinfo {author} {\bibfnamefont {J.}~\bibnamefont
  {Dudek}},\ }\href {\doibase 10.1103/PhysRevC.56.1795} {\bibfield  {journal}
  {\bibinfo  {journal} {Phys. Rev. C}\ }\textbf {\bibinfo {volume} {56}},\
  \bibinfo {pages} {1795} (\bibinfo {year} {1997})}\BibitemShut {NoStop}%
\bibitem [{\citenamefont {Volya}\ \emph {et~al.}(2001)\citenamefont {Volya},
  \citenamefont {Brown},\ and\ \citenamefont {Zelevinsky}}]{Volya:2000ne}%
  \BibitemOpen
  \bibfield  {author} {\bibinfo {author} {\bibfnamefont {A.}~\bibnamefont
  {Volya}}, \bibinfo {author} {\bibfnamefont {B.~A.}\ \bibnamefont {Brown}}, \
  and\ \bibinfo {author} {\bibfnamefont {V.}~\bibnamefont {Zelevinsky}},\
  }\href {\doibase 10.1016/S0370-2693(01)00431-2} {\bibfield  {journal}
  {\bibinfo  {journal} {Phys. Lett. B}\ }\textbf {\bibinfo {volume} {509}},\
  \bibinfo {pages} {37} (\bibinfo {year} {2001})},\ \Eprint
  {http://arxiv.org/abs/nucl-th/0011079} {arXiv:nucl-th/0011079} \BibitemShut
  {NoStop}%
\bibitem [{\citenamefont {Zelevinsky}\ and\ \citenamefont
  {Volya}(2003)}]{Zelevinsky:2003ad}%
  \BibitemOpen
  \bibfield  {author} {\bibinfo {author} {\bibfnamefont {V.}~\bibnamefont
  {Zelevinsky}}\ and\ \bibinfo {author} {\bibfnamefont {A.}~\bibnamefont
  {Volya}},\ }\href {\doibase 10.1134/1.1619492} {\bibfield  {journal}
  {\bibinfo  {journal} {Phys. Atom. Nucl.}\ }\textbf {\bibinfo {volume} {66}},\
  \bibinfo {pages} {1781} (\bibinfo {year} {2003})},\ \Eprint
  {http://arxiv.org/abs/nucl-th/0303010} {arXiv:nucl-th/0303010} \BibitemShut
  {NoStop}%
\bibitem [{\citenamefont {Liu}\ and\ \citenamefont {Qi}(2021)}]{Liu:2020mkp}%
  \BibitemOpen
  \bibfield  {author} {\bibinfo {author} {\bibfnamefont {X.}~\bibnamefont
  {Liu}}\ and\ \bibinfo {author} {\bibfnamefont {C.}~\bibnamefont {Qi}},\
  }\href {\doibase 10.1016/j.cpc.2020.107349} {\bibfield  {journal} {\bibinfo
  {journal} {Comput. Phys. Commun.}\ }\textbf {\bibinfo {volume} {259}},\
  \bibinfo {pages} {107349} (\bibinfo {year} {2021})},\ \Eprint
  {http://arxiv.org/abs/2001.01978} {arXiv:2001.01978 [nucl-th]} \BibitemShut
  {NoStop}%
\bibitem [{\citenamefont {Sandulescu}\ \emph {et~al.}(1997)\citenamefont
  {Sandulescu}, \citenamefont {Blomqvist}, \citenamefont {Engeland},
  \citenamefont {Hjorth-Jensen}, \citenamefont {Holt}, \citenamefont {Liotta},\
  and\ \citenamefont {Osnes}}]{Sandulescu:1996wq}%
  \BibitemOpen
  \bibfield  {author} {\bibinfo {author} {\bibfnamefont {N.}~\bibnamefont
  {Sandulescu}}, \bibinfo {author} {\bibfnamefont {J.}~\bibnamefont
  {Blomqvist}}, \bibinfo {author} {\bibfnamefont {T.}~\bibnamefont {Engeland}},
  \bibinfo {author} {\bibfnamefont {M.}~\bibnamefont {Hjorth-Jensen}}, \bibinfo
  {author} {\bibfnamefont {A.}~\bibnamefont {Holt}}, \bibinfo {author}
  {\bibfnamefont {R.~J.}\ \bibnamefont {Liotta}}, \ and\ \bibinfo {author}
  {\bibfnamefont {E.}~\bibnamefont {Osnes}},\ }\href {\doibase
  10.1103/PhysRevC.55.2708} {\bibfield  {journal} {\bibinfo  {journal} {Phys.
  Rev. C}\ }\textbf {\bibinfo {volume} {55}},\ \bibinfo {pages} {2708}
  (\bibinfo {year} {1997})},\ \Eprint {http://arxiv.org/abs/nucl-th/9612052}
  {arXiv:nucl-th/9612052} \BibitemShut {NoStop}%
\bibitem [{\citenamefont {Jia}(2013)}]{Jia:2013lea}%
  \BibitemOpen
  \bibfield  {author} {\bibinfo {author} {\bibfnamefont {L.~Y.}\ \bibnamefont
  {Jia}},\ }\href {\doibase 10.1103/PhysRevC.88.044303} {\bibfield  {journal}
  {\bibinfo  {journal} {Phys. Rev. C}\ }\textbf {\bibinfo {volume} {88}},\
  \bibinfo {pages} {044303} (\bibinfo {year} {2013})},\ \Eprint
  {http://arxiv.org/abs/1305.2697} {arXiv:1305.2697 [nucl-th]} \BibitemShut
  {NoStop}%
\bibitem [{\citenamefont
  {{Richardson}}(1963{\natexlab{a}})}]{Richardson:1963a}%
  \BibitemOpen
  \bibfield  {author} {\bibinfo {author} {\bibfnamefont {R.~W.}\ \bibnamefont
  {{Richardson}}},\ }\href {\doibase 10.1016/0031-9163(63)90259-2} {\bibfield
  {journal} {\bibinfo  {journal} {Physics Letters}\ }\textbf {\bibinfo {volume}
  {3}},\ \bibinfo {pages} {277} (\bibinfo {year}
  {1963}{\natexlab{a}})}\BibitemShut {NoStop}%
\bibitem [{\citenamefont
  {{Richardson}}(1963{\natexlab{b}})}]{Richardson:1963b}%
  \BibitemOpen
  \bibfield  {author} {\bibinfo {author} {\bibfnamefont {R.~W.}\ \bibnamefont
  {{Richardson}}},\ }\href {\doibase 10.1016/S0375-9601(63)80039-0} {\bibfield
  {journal} {\bibinfo  {journal} {Physics Letters}\ }\textbf {\bibinfo {volume}
  {5}},\ \bibinfo {pages} {82} (\bibinfo {year}
  {1963}{\natexlab{b}})}\BibitemShut {NoStop}%
\bibitem [{\citenamefont {{Gaudin, M.}}(1976)}]{Gaudin:1976}%
  \BibitemOpen
  \bibfield  {author} {\bibinfo {author} {\bibnamefont {{Gaudin, M.}}},\ }\href
  {\doibase 10.1051/jphys:0197600370100108700} {\bibfield  {journal} {\bibinfo
  {journal} {J. Phys. France}\ }\textbf {\bibinfo {volume} {37}},\ \bibinfo
  {pages} {1087} (\bibinfo {year} {1976})}\BibitemShut {NoStop}%
\bibitem [{\citenamefont {Dukelsky}\ \emph {et~al.}(2011)\citenamefont
  {Dukelsky}, \citenamefont {Lerma~H.}, \citenamefont {Robledo}, \citenamefont
  {Rodriguez-Guzman},\ and\ \citenamefont {Rombouts}}]{Dukelsky:2011nb}%
  \BibitemOpen
  \bibfield  {author} {\bibinfo {author} {\bibfnamefont {J.}~\bibnamefont
  {Dukelsky}}, \bibinfo {author} {\bibfnamefont {S.}~\bibnamefont {Lerma~H.}},
  \bibinfo {author} {\bibfnamefont {L.~M.}\ \bibnamefont {Robledo}}, \bibinfo
  {author} {\bibfnamefont {R.}~\bibnamefont {Rodriguez-Guzman}}, \ and\
  \bibinfo {author} {\bibfnamefont {S.~M.~A.}\ \bibnamefont {Rombouts}},\
  }\href {\doibase 10.1103/PhysRevC.84.061301} {\bibfield  {journal} {\bibinfo
  {journal} {Phys. Rev. C}\ }\textbf {\bibinfo {volume} {84}},\ \bibinfo
  {pages} {061301} (\bibinfo {year} {2011})},\ \Eprint
  {http://arxiv.org/abs/1109.4292} {arXiv:1109.4292 [nucl-th]} \BibitemShut
  {NoStop}%
\bibitem [{\citenamefont {{Guan}}\ and\ \citenamefont
  {{Qi}}(2022)}]{Guan:2022CoPhC}%
  \BibitemOpen
  \bibfield  {author} {\bibinfo {author} {\bibfnamefont {X.}~\bibnamefont
  {{Guan}}}\ and\ \bibinfo {author} {\bibfnamefont {C.}~\bibnamefont {{Qi}}},\
  }\href {\doibase 10.1016/j.cpc.2022.108310} {\bibfield  {journal} {\bibinfo
  {journal} {Computer Physics Communications}\ }\textbf {\bibinfo {volume}
  {275}},\ \bibinfo {eid} {108310} (\bibinfo {year} {2022})}\BibitemShut
  {NoStop}%
\bibitem [{\citenamefont {Rombouts}\ \emph {et~al.}(2004)\citenamefont
  {Rombouts}, \citenamefont {Van~Neck},\ and\ \citenamefont
  {Dukelsky}}]{Rombouts:2003zd}%
  \BibitemOpen
  \bibfield  {author} {\bibinfo {author} {\bibfnamefont {S.}~\bibnamefont
  {Rombouts}}, \bibinfo {author} {\bibfnamefont {D.}~\bibnamefont {Van~Neck}},
  \ and\ \bibinfo {author} {\bibfnamefont {J.}~\bibnamefont {Dukelsky}},\
  }\href {\doibase 10.1103/PhysRevC.69.061303} {\bibfield  {journal} {\bibinfo
  {journal} {Phys. Rev. C}\ }\textbf {\bibinfo {volume} {69}},\ \bibinfo
  {pages} {061303} (\bibinfo {year} {2004})},\ \Eprint
  {http://arxiv.org/abs/nucl-th/0312070} {arXiv:nucl-th/0312070} \BibitemShut
  {NoStop}%
\bibitem [{\citenamefont {{Metropolis}}\ \emph {et~al.}(1953)\citenamefont
  {{Metropolis}}, \citenamefont {{Rosenbluth}}, \citenamefont {{Rosenbluth}},
  \citenamefont {{Teller}},\ and\ \citenamefont {{Teller}}}]{Metropolis:1953}%
  \BibitemOpen
  \bibfield  {author} {\bibinfo {author} {\bibfnamefont {N.}~\bibnamefont
  {{Metropolis}}}, \bibinfo {author} {\bibfnamefont {A.~W.}\ \bibnamefont
  {{Rosenbluth}}}, \bibinfo {author} {\bibfnamefont {M.~N.}\ \bibnamefont
  {{Rosenbluth}}}, \bibinfo {author} {\bibfnamefont {A.~H.}\ \bibnamefont
  {{Teller}}}, \ and\ \bibinfo {author} {\bibfnamefont {E.}~\bibnamefont
  {{Teller}}},\ }\href {\doibase 10.1063/1.1699114} {\bibfield  {journal}
  {\bibinfo  {journal} {\jcp}\ }\textbf {\bibinfo {volume} {21}},\ \bibinfo
  {pages} {1087} (\bibinfo {year} {1953})}\BibitemShut {NoStop}%
\bibitem [{\citenamefont {Hastings}(1970)}]{Hastings:1970}%
  \BibitemOpen
  \bibfield  {author} {\bibinfo {author} {\bibfnamefont {W.~K.}\ \bibnamefont
  {Hastings}},\ }\href {\doibase 10.1093/biomet/57.1.97} {\bibfield  {journal}
  {\bibinfo  {journal} {Biometrika}\ }\textbf {\bibinfo {volume} {57}},\
  \bibinfo {pages} {97} (\bibinfo {year} {1970})},\ \Eprint
  {http://arxiv.org/abs/https://academic.oup.com/biomet/article-pdf/57/1/97/23940249/57-1-97.pdf}
  {https://academic.oup.com/biomet/article-pdf/57/1/97/23940249/57-1-97.pdf}
  \BibitemShut {NoStop}%
\bibitem [{\citenamefont {Sorella}(2005)}]{Sorella:2005}%
  \BibitemOpen
  \bibfield  {author} {\bibinfo {author} {\bibfnamefont {S.}~\bibnamefont
  {Sorella}},\ }\href {\doibase 10.1103/PhysRevB.71.241103} {\bibfield
  {journal} {\bibinfo  {journal} {Phys. Rev. B}\ }\textbf {\bibinfo {volume}
  {71}},\ \bibinfo {pages} {241103} (\bibinfo {year} {2005})}\BibitemShut
  {NoStop}%
\bibitem [{\citenamefont {Amari}(1998)}]{amari_natural_1998}%
  \BibitemOpen
  \bibfield  {author} {\bibinfo {author} {\bibfnamefont {S.-i.}\ \bibnamefont
  {Amari}},\ }\href {\doibase 10.1162/089976698300017746} {\bibfield  {journal}
  {\bibinfo  {journal} {Neural Computation}\ }\textbf {\bibinfo {volume}
  {10}},\ \bibinfo {pages} {251} (\bibinfo {year} {1998})}\BibitemShut
  {NoStop}%
\bibitem [{\citenamefont {{Stokes}}\ \emph {et~al.}(2019)\citenamefont
  {{Stokes}}, \citenamefont {{Izaac}}, \citenamefont {{Killoran}},\ and\
  \citenamefont {{Carleo}}}]{Stokes:2019}%
  \BibitemOpen
  \bibfield  {author} {\bibinfo {author} {\bibfnamefont {J.}~\bibnamefont
  {{Stokes}}}, \bibinfo {author} {\bibfnamefont {J.}~\bibnamefont {{Izaac}}},
  \bibinfo {author} {\bibfnamefont {N.}~\bibnamefont {{Killoran}}}, \ and\
  \bibinfo {author} {\bibfnamefont {G.}~\bibnamefont {{Carleo}}},\ }\href@noop
  {} {\bibfield  {journal} {\bibinfo  {journal} {arXiv e-prints}\ ,\ \bibinfo
  {eid} {arXiv:1909.02108}} (\bibinfo {year} {2019})},\ \Eprint
  {http://arxiv.org/abs/1909.02108} {arXiv:1909.02108 [quant-ph]} \BibitemShut
  {NoStop}%
\bibitem [{\citenamefont {Sorella}(1998)}]{sorella_green_1998}%
  \BibitemOpen
  \bibfield  {author} {\bibinfo {author} {\bibfnamefont {S.}~\bibnamefont
  {Sorella}},\ }\href {\doibase 10.1103/PhysRevLett.80.4558} {\bibfield
  {journal} {\bibinfo  {journal} {Physical Review Letters}\ }\textbf {\bibinfo
  {volume} {80}},\ \bibinfo {pages} {4558} (\bibinfo {year}
  {1998})}\BibitemShut {NoStop}%
\bibitem [{\citenamefont {Neuscamman}\ \emph {et~al.}(2012)\citenamefont
  {Neuscamman}, \citenamefont {Umrigar},\ and\ \citenamefont
  {Chan}}]{Neuscamman:2012}%
  \BibitemOpen
  \bibfield  {author} {\bibinfo {author} {\bibfnamefont {E.}~\bibnamefont
  {Neuscamman}}, \bibinfo {author} {\bibfnamefont {C.~J.}\ \bibnamefont
  {Umrigar}}, \ and\ \bibinfo {author} {\bibfnamefont {G.~K.-L.}\ \bibnamefont
  {Chan}},\ }\href {\doibase 10.1103/PhysRevB.85.045103} {\bibfield  {journal}
  {\bibinfo  {journal} {Phys. Rev. B}\ }\textbf {\bibinfo {volume} {85}},\
  \bibinfo {pages} {045103} (\bibinfo {year} {2012})}\BibitemShut {NoStop}%
\bibitem [{\citenamefont {Tijmen~Tieleman}(2018)}]{Hinton:2018}%
  \BibitemOpen
  \bibfield  {author} {\bibinfo {author} {\bibfnamefont {G.~H.}\ \bibnamefont
  {Tijmen~Tieleman}},\ }\href
  {https://www.cs.toronto.edu/~tijmen/csc321/slides/lecture_slides_lec6.pdf}
  {\enquote {\bibinfo {title} {Lecture 6.5 - {RMSProp}},}\ }\bibinfo
  {howpublished} {Coursera Neural Networks for Machine Learning lecture 6}
  (\bibinfo {year} {2018})\BibitemShut {NoStop}%
\bibitem [{\citenamefont {{Kingma}}\ and\ \citenamefont
  {{Ba}}(2014)}]{Kingma:2014}%
  \BibitemOpen
  \bibfield  {author} {\bibinfo {author} {\bibfnamefont {D.~P.}\ \bibnamefont
  {{Kingma}}}\ and\ \bibinfo {author} {\bibfnamefont {J.}~\bibnamefont
  {{Ba}}},\ }\href@noop {} {\bibfield  {journal} {\bibinfo  {journal} {arXiv
  e-prints}\ ,\ \bibinfo {eid} {arXiv:1412.6980}} (\bibinfo {year} {2014})},\
  \Eprint {http://arxiv.org/abs/1412.6980} {arXiv:1412.6980 [cs.LG]}
  \BibitemShut {NoStop}%
\bibitem [{\citenamefont {{Coester}}(1958)}]{Coester:1958}%
  \BibitemOpen
  \bibfield  {author} {\bibinfo {author} {\bibfnamefont {F.}~\bibnamefont
  {{Coester}}},\ }\href {\doibase 10.1016/0029-5582(58)90280-3} {\bibfield
  {journal} {\bibinfo  {journal} {Nuclear Physics}\ }\textbf {\bibinfo {volume}
  {7}},\ \bibinfo {pages} {421} (\bibinfo {year} {1958})}\BibitemShut {NoStop}%
\bibitem [{\citenamefont {{Coester}}\ and\ \citenamefont
  {{K{\"u}mmel}}(1960)}]{Coester:1960}%
  \BibitemOpen
  \bibfield  {author} {\bibinfo {author} {\bibfnamefont {F.}~\bibnamefont
  {{Coester}}}\ and\ \bibinfo {author} {\bibfnamefont {H.}~\bibnamefont
  {{K{\"u}mmel}}},\ }\href {\doibase 10.1016/0029-5582(60)90140-1} {\bibfield
  {journal} {\bibinfo  {journal} {Nuclear Physics}\ }\textbf {\bibinfo {volume}
  {17}},\ \bibinfo {pages} {477} (\bibinfo {year} {1960})}\BibitemShut
  {NoStop}%
\bibitem [{\citenamefont {Wick}(1950)}]{Wick1950}%
  \BibitemOpen
  \bibfield  {author} {\bibinfo {author} {\bibfnamefont {G.~C.}\ \bibnamefont
  {Wick}},\ }\href {\doibase 10.1103/PhysRev.80.268} {\bibfield  {journal}
  {\bibinfo  {journal} {Phys. Rev.}\ }\textbf {\bibinfo {volume} {80}},\
  \bibinfo {pages} {268} (\bibinfo {year} {1950})}\BibitemShut {NoStop}%
\bibitem [{\citenamefont {Paldus}(2006)}]{Paldus2006}%
  \BibitemOpen
  \bibfield  {author} {\bibinfo {author} {\bibfnamefont {J.}~\bibnamefont
  {Paldus}},\ }\href {\doibase 10.1007/978-0-387-26308-3_5} {\emph {\bibinfo
  {title} {Springer Handbook of Atomic, Molecular, and Optical Physics}}},\
  edited by\ \bibinfo {editor} {\bibfnamefont {G.}~\bibnamefont {Drake}}\
  (\bibinfo  {publisher} {Springer New York},\ \bibinfo {address} {New York,
  NY},\ \bibinfo {year} {2006})\ pp.\ \bibinfo {pages} {101--114}\BibitemShut
  {NoStop}%
\bibitem [{\citenamefont {Shavitt}\ and\ \citenamefont
  {Bartlett}(2009)}]{shavittbartlett2009}%
  \BibitemOpen
  \bibfield  {author} {\bibinfo {author} {\bibfnamefont {I.}~\bibnamefont
  {Shavitt}}\ and\ \bibinfo {author} {\bibfnamefont {R.~J.}\ \bibnamefont
  {Bartlett}},\ }\href {\doibase 10.1017/CBO9780511596834} {\emph {\bibinfo
  {title} {Many-Body Methods in Chemistry and Physics: MBPT and Coupled-Cluster
  Theory}}},\ Cambridge Molecular Science\ (\bibinfo  {publisher} {Cambridge
  University Press},\ \bibinfo {year} {2009})\BibitemShut {NoStop}%
\bibitem [{\citenamefont {Lietz}\ \emph {et~al.}(2017)\citenamefont {Lietz},
  \citenamefont {Jansen}, \citenamefont {Hagen},\ and\ \citenamefont
  {Hjorth-Jensen}}]{morten_book}%
  \BibitemOpen
  \bibfield  {author} {\bibinfo {author} {\bibfnamefont {J.}~\bibnamefont
  {Lietz}}, \bibinfo {author} {\bibfnamefont {G.}~\bibnamefont {Jansen}},
  \bibinfo {author} {\bibfnamefont {G.}~\bibnamefont {Hagen}}, \ and\ \bibinfo
  {author} {\bibfnamefont {M.}~\bibnamefont {Hjorth-Jensen}},\ }\enquote
  {\bibinfo {title} {Computational nucl. phys. and post hartree-fock
  methods},}\ in\ \href@noop {} {\emph {\bibinfo {booktitle} {An advanced
  course in computational Nucl. Phys. bridging the scales from quarks to
  neutron stars}}}\ (\bibinfo  {publisher} {Springer International
  Publishing},\ \bibinfo {year} {2017})\ p.\ \bibinfo {pages}
  {327–333}\BibitemShut {NoStop}%
\bibitem [{\citenamefont {{Ortiz}}\ \emph {et~al.}(2014)\citenamefont
  {{Ortiz}}, \citenamefont {{Dukelsky}}, \citenamefont {{Cobanera}},
  \citenamefont {{Esebbag}},\ and\ \citenamefont
  {{Beenakker}}}]{Ortiz:2014PhRvL}%
  \BibitemOpen
  \bibfield  {author} {\bibinfo {author} {\bibfnamefont {G.}~\bibnamefont
  {{Ortiz}}}, \bibinfo {author} {\bibfnamefont {J.}~\bibnamefont {{Dukelsky}}},
  \bibinfo {author} {\bibfnamefont {E.}~\bibnamefont {{Cobanera}}}, \bibinfo
  {author} {\bibfnamefont {C.}~\bibnamefont {{Esebbag}}}, \ and\ \bibinfo
  {author} {\bibfnamefont {C.}~\bibnamefont {{Beenakker}}},\ }\href {\doibase
  10.1103/PhysRevLett.113.267002} {\bibfield  {journal} {\bibinfo  {journal}
  {\prl}\ }\textbf {\bibinfo {volume} {113}},\ \bibinfo {eid} {267002}
  (\bibinfo {year} {2014})}\BibitemShut {NoStop}%
\bibitem [{\citenamefont {{Iemini}}\ \emph {et~al.}(2015)\citenamefont
  {{Iemini}}, \citenamefont {{Mazza}}, \citenamefont {{Rossini}}, \citenamefont
  {{Fazio}},\ and\ \citenamefont {{Diehl}}}]{Iemini:2015PhRvL}%
  \BibitemOpen
  \bibfield  {author} {\bibinfo {author} {\bibfnamefont {F.}~\bibnamefont
  {{Iemini}}}, \bibinfo {author} {\bibfnamefont {L.}~\bibnamefont {{Mazza}}},
  \bibinfo {author} {\bibfnamefont {D.}~\bibnamefont {{Rossini}}}, \bibinfo
  {author} {\bibfnamefont {R.}~\bibnamefont {{Fazio}}}, \ and\ \bibinfo
  {author} {\bibfnamefont {S.}~\bibnamefont {{Diehl}}},\ }\href {\doibase
  10.1103/PhysRevLett.115.156402} {\bibfield  {journal} {\bibinfo  {journal}
  {\prl}\ }\textbf {\bibinfo {volume} {115}},\ \bibinfo {eid} {156402}
  (\bibinfo {year} {2015})},\ \Eprint {http://arxiv.org/abs/1504.04230}
  {arXiv:1504.04230 [cond-mat.str-el]} \BibitemShut {NoStop}%
\bibitem [{\citenamefont {Hjorth-Jensen}\ \emph {et~al.}(1995)\citenamefont
  {Hjorth-Jensen}, \citenamefont {Kuo},\ and\ \citenamefont {Osnes}}]{hko95}%
  \BibitemOpen
  \bibfield  {author} {\bibinfo {author} {\bibfnamefont {M.}~\bibnamefont
  {Hjorth-Jensen}}, \bibinfo {author} {\bibfnamefont {T.~T.}\ \bibnamefont
  {Kuo}}, \ and\ \bibinfo {author} {\bibfnamefont {E.}~\bibnamefont {Osnes}},\
  }\href {\doibase https://doi.org/10.1016/0370-1573(95)00012-6} {\bibfield
  {journal} {\bibinfo  {journal} {Physics Reports}\ }\textbf {\bibinfo {volume}
  {261}},\ \bibinfo {pages} {125} (\bibinfo {year} {1995})}\BibitemShut
  {NoStop}%
\bibitem [{\citenamefont {Tichai}\ \emph {et~al.}(2022)\citenamefont {Tichai},
  \citenamefont {Knecht}, \citenamefont {Kruppa}, \citenamefont {Legeza},
  \citenamefont {Moca}, \citenamefont {Schwenk}, \citenamefont {Werner},\ and\
  \citenamefont {Zarand}}]{Tichai:2022bxr}%
  \BibitemOpen
  \bibfield  {author} {\bibinfo {author} {\bibfnamefont {A.}~\bibnamefont
  {Tichai}}, \bibinfo {author} {\bibfnamefont {S.}~\bibnamefont {Knecht}},
  \bibinfo {author} {\bibfnamefont {A.~T.}\ \bibnamefont {Kruppa}}, \bibinfo
  {author} {\bibfnamefont {O.}~\bibnamefont {Legeza}}, \bibinfo {author}
  {\bibfnamefont {C.~P.}\ \bibnamefont {Moca}}, \bibinfo {author}
  {\bibfnamefont {A.}~\bibnamefont {Schwenk}}, \bibinfo {author} {\bibfnamefont
  {M.~A.}\ \bibnamefont {Werner}}, \ and\ \bibinfo {author} {\bibfnamefont
  {G.}~\bibnamefont {Zarand}},\ }\href@noop {} {\  (\bibinfo {year} {2022})},\
  \Eprint {http://arxiv.org/abs/2207.01438} {arXiv:2207.01438 [nucl-th]}
  \BibitemShut {NoStop}%
\end{thebibliography}%
\end{document}